\documentclass[pdflatex,sn-mathphys-num]{sn-jnl}% Math and Physical Sciences Numbered Reference Style 
\usepackage{graphicx}%
\usepackage{multirow}%
\usepackage{amsmath,amssymb,amsfonts}%
\usepackage{amsthm}%
\usepackage{mathrsfs}%
\usepackage[title]{appendix}%
\usepackage{xcolor}%
\usepackage{textcomp}%
\usepackage{manyfoot}%
\usepackage{booktabs}%
\usepackage{siunitx}
\usepackage{caption}
\usepackage{algorithm}%
\usepackage{algorithmicx}%
\usepackage{algpseudocode}%
\usepackage{listings}%
\usepackage{chngcntr}
\usepackage{float}
\usepackage{pdflscape}
\usepackage{verbatim}
\counterwithout{figure}{section}

\graphicspath{ {images/} }
\usepackage{lineno}
\theoremstyle{thmstyleone}%
\theoremstyle{thmstyletwo}%

\theoremstyle{thmstylethree}%

\begin{document}
%\detailtexcount{sn-article}
\title[Article Title]{Airborne assessment uncovers socioeconomic stratification of urban nature in England}
%Option 2:Beyond the Canopy: Deconstructing the 3-30-300 rule reveals new geographies of environmental injustice 
%Option 3:Cracking the Code of a Green City: A computational framework for the 3-30-300 rule uncovers deep environmental inequality in England

%{Geospatial Analysis Highlights North-South Division in Access to Nature in Cities}
%\North/South divide In England in Nature availability 
%\Nature inequality in England regions demonstrated by the 3-30-300 rule and geospatial analysis
%\Do people thrive without nature?

\author*[1,2]{\fnm{Andrés Camilo} \sur{Zúñiga-González}}\email{acz25@cam.ac.uk}

\author[2]{\fnm{Anil} \sur{Madhavapeddy}}\email{avsm2@cam.ac.uk}
\equalcont{These authors contributed equally to this work.}

\author[1]{\fnm{Ronita} \sur{Bardhan}}\email{rb867@cam.ac.uk}
\equalcont{These authors contributed equally to this work.}

\affil*[1]{\orgdiv{Department of Architecture}, \orgname{University of Cambridge}, \orgaddress{\street{Street}, \city{Cambridge}, \postcode{100190}, \state{Cambridgeshire}, \country{United Kingdom}}}

\affil[2]{\orgdiv{Department of Computer Science and Technology}, \orgname{University of Cambridge}, \orgaddress{\street{Street}, \city{Cambridge}, \postcode{100190}, \state{Cambridgeshire}, \country{United Kingdom}}}

\abstract{Nature access is increasingly recognised as a public health and equity imperative, yet cities lack standardised ways to measure who benefits from green infrastructure. We present the first national, building-level assessment of the 3–30–300 urban greening rule across England, integrating high-performance computing with open LiDAR and geospatial datasets. Our framework quantifies proximity, availability, and accessibility of greenery, linking each to socioeconomic deprivation through Gini-based inequality metrics. Results reveal that while most English residents meet the tree visibility criterion, only 0.1\% of urban areas meet all three thresholds; canopy cover and park access sharply diverge along deprivation lines. Wealthier areas enjoy greater ambient greenness, whereas deprived urban cores often have better proximity to parks but lower vegetation density. The study offers a scalable computational blueprint for assessing nature equity and demonstrates that green accessibility represents a new dimension of socioeconomic inequality. These findings call for policy approaches that move beyond proximity metrics toward equitable, quality-based standards for nature access.}


\maketitle

\section{Introduction} \label{sec:introduction}

Urban green infrastructure is crucial for urban resilience and public health, providing a vital nature-based solution to the pressures posed by climate change and urbanisation. This becomes a crucial factor in ever-growing cities, since approximately 70\% of the global population is projected to live in urban areas~\cite{united_nations_department_of_economic_and_social_affairs_population_division_world_2019}. Multiple studies have highlighted the role of urban greenery in mitigating air pollution, reducing urban heat island effects (higher temperatures in urban areas due absorption of heat by built-up surfaces), and fostering social cohesion \cite{kowarik_promoting_2025}. Moreover, accessible green spaces are linked to improved physical and mental well-being, positioning them as a pillar of sustainable urban planning with human health as a focus \cite{giannico_green_2021}.

Effective urban greening policies depend on reliable and consistent measurement, yet no standard process exists. Methodologies, spatial scales, and time frames for analysis differ widely between studies, making it impossible to compare cities accurately or establish global benchmarks. This inconsistency is particularly damaging for research that connects green infrastructure to social and health outcomes, as the lack of a standard metric for ``greenness'' undermines the findings. A standardised method is therefore crucial for building effective, evidence-based greening policies worldwide~\cite{browning_measuring_2024, zhang_associations_2019, battiston_need_2024}.

Existing efforts to quantify urban green infrastructure are typically conducted at two distinct scales: granular analyses within individual cities \cite{juergens_experimental_2022, boehnke_mapping_2022, derkzen_review_2015, seidl_green_2021} or broad continental comparisons. The latter have revealed large-scale patterns, such as a North-South divide in green space accessibility across Europe, where northern cities are generally greener \cite{kabisch_urban_2016}. Zooming into the British context, research has confirmed a strong link between urban form and inequality, finding that populous, deprived urban centres tend to have less canopy cover \cite{robinson_urban_2022}. What remains absent, however, is a consistent, high-resolution analysis at the national scale capable of moving beyond city-specific findings or regional averages to systematically map the nuanced geography of environmental inequality.

In response to this need for simplified and actionable metrics, the ``3-30-300 rule'' has recently been proposed and is gaining significant traction in urban policy circles around Europe and North America~\cite{konijnendijk_evidence-based_2023}. This rule of thumb offers an intuitive framework for what a sufficiently green city should look like: every citizen should see at least 3 trees from their home, school, and workplace; every neighbourhood should have at least 30\% canopy cover; and every resident should live within 300 metres of a quality public green space. Together, these components aim to quantify the visibility, availability, and accessibility of urban nature, respectively \cite{zheng_quantitative_2024}.

Despite its growing adoption in cities such as Singapore, Amsterdam (Netherlands), Buenos Aires (Argentina), Sydney, Melbourne (Australia), New York, Seattle, Denver (USA)\cite{nieuwenhuijsen_evaluation_2022, croeser_acute_2024} and in the entire Italian territory\cite{giannico_mortality_2024}, the 3-30-300 rule currently lacks a standardised methodological foundation for assessment at scale. Current approaches are fragmented, ranging from satellite-derived vegetation indices, such as NDVI, to labour-intensive street-level photograph analysis \cite{zheng_quantitative_2024, browning_measuring_2024}, and even survey-based methods. This methodological void prevents a systematic and comparable evaluation and makes it difficult to ascertain whether the rule truly captures the nuances of nature exposure and accessibility.

Due to the lack of a standard methodology, we develop and apply a novel computational framework to conduct the first national-scale, building-level assessment of the 3-30-300 rule, using England as our case study. The methodology presented applies high-performance computing on publicly available LiDAR and geospatial data to overcome previous limitations of scale and resolution. Specifically, we: 1) develop and validate novel, high-resolution proxies for the rule. For the ``3 visible trees'' component, we move beyond simple tree counts to a regression-based metric that captures the density of the surrounding tree canopy from each building's perspective. For the '300 m to a park' component, we use a detailed national road network to model realistic walking distances, a significant improvement over simpler distance-based estimations; 2) directly quantify environmental inequality by calculating Gini coefficients \cite{martin_using_2025} from this building-level data on residential units, revealing disparities within local areas and regions; and 3) critically evaluate how these human-centric 3-30-300 metrics correlate and integrate with traditional remote-sensing-derived indices.

By providing a methodological framework, this research facilitates the standardised application of the 3-30-300 rule in other countries with comparable data. Ultimately, we aim to determine if this increasingly influential rule, when combined with remote sensing data, provides a robust framework for guiding evidence-based urban planning and promoting equitable access to urban nature in other countries. Finally, our results highlight that most English citizens live in areas where the 30 and 300 components thresholds are not met, while the 3-component is largely fulfilled by most areas. When comparing to deprivation levels, the tree proximity and canopy cover metrics have higher attainment in less deprived areas, but public parks are more accessible in more deprived areas. Furthermore, we provide an inequality analysis that highlights a nature deprivation difference by regions, revealing that disparities in nature access align with and potentially exacerbate existing inter and intra-regional socio-economic inequalities, indicating that nature accessibility is a new layer of inequality to be considered.

\section{Results} \label{sec:results}

\subsection{How many trees are there in English cities?} \label{subsec:how_many_trees}

Our national LiDAR segmentation identified approximately 190 million trees across England. A key finding is the stark urban-rural divide: only 26.9\% of these trees are located within urban areas as defined by the ONS. For the subsequent analysis of the ``3-30-300'' rule, we used a filtered dataset of ~156 million significant trees (those with a height \textgreater 3 m and crown area \textgreater 10 m²; see Methods). This entire analysis was performed across 32,742 Lower Layer Super Output Areas (LSOAs), which represent the geographies for which consistent 2019 Index of Multiple Deprivation (IMD) data were available, explaining minor discrepancies with the total number of official 2021 LSOAs. A map of all the areas covered by the segmentation algorithm is shown in Figure \ref{fig:data_coverage_map}.

\subsection{Tree Distribution: Per Capita vs. Absolute Density} \label{subsec:tree_distribution}

\begin{figure}[H] 
\centering
\includegraphics[width=\textwidth]{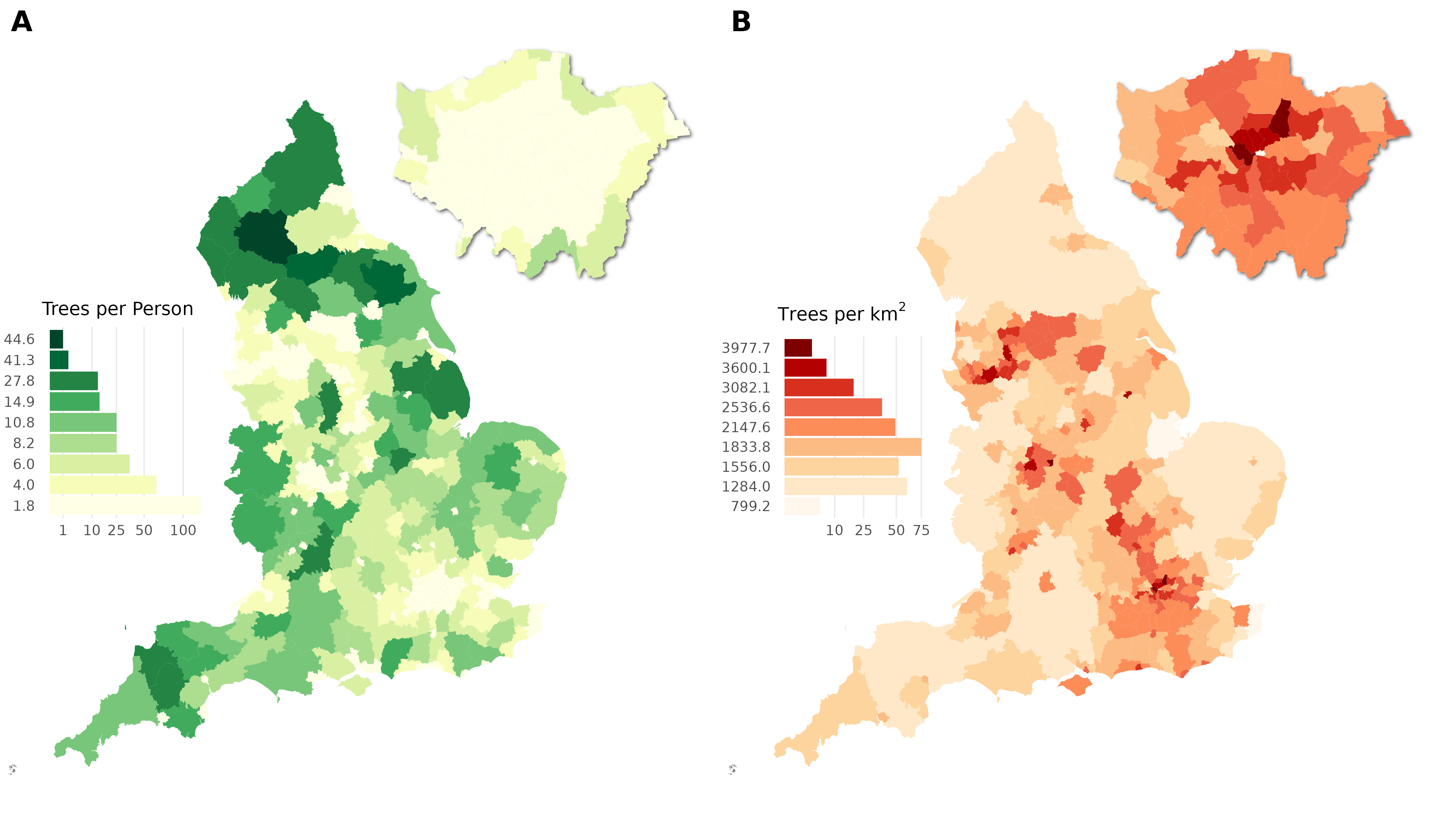}
\caption{Contrasting spatial distributions of per capita and absolute tree density in England. Choropleth maps illustrate two metrics of tree distribution at the Local Authority District level. \textbf{(A)} Trees per person, a measure of per capita tree availability. \textbf{(B)} Trees per km², a measure of absolute tree density. Map insets depict the Greater London area.}
    
\label{fig:total_tree_distribution}
\end{figure}

Our analysis of tree distribution across England reveals a fundamental dichotomy: the greenest parts of the country can be identified as either its most rural expanses or its most densely populated urban regions, depending entirely on the metric used. This finding (Figure 1) presents a critical measurement contradiction for urban planners and policymakers, as the choice of metric can lead to dramatically different conclusions about where greening interventions are most needed (Figure \ref{fig:total_tree_distribution}).

When viewed through a per capita lens—a measure of how many trees are available per person—resources appear most plentiful in rural, sparsely populated regions. Local Authority Districts (LADs) in the North of England (e.g., Northumberland, Cumbria) and the South West exhibit the highest values, often exceeding 40 trees per person (Figure \ref{fig:total_tree_distribution}A). From this perspective, a clear urban-rural divide emerges, suggesting that residents of major metropolitan areas like London and Manchester have the least access to tree resources.

However, this narrative is completely inverted when considering absolute density (trees per km²), which measures the concentration of trees in a given area (Figure \ref{fig:total_tree_distribution}B). Here, the highest concentrations are found not in the countryside, but in the densely populated and affluent South East. The commuter belt surrounding Greater London, along with parts of London itself, shows the highest densities, with many areas containing over 3,000 trees per km². This finding challenges the simplistic view of cities as concrete jungles, revealing that many urban and suburban areas contain an exceptionally high overall stock of trees, even if they are shared among more people.

\subsection{Quantification and Attainment of the 3-30-300}
\label{subsec:quanti_3-30-300}
When accounting for the areas with high access to trees and other forms of green infrastructure, we only considered those in LSOAs classified as urban by the ONS, which further reduced the number of areas, as seen in Table \ref{tab:3-30-300_table}. Most areas complied with having an average of more than 3 visible trees with no significant differences between regions, while canopy cover presented the lowest percentage of high cover, with the South West considerably outclassing the other regions. Moreover, access to parks was significantly higher in London than in other regions, with the Eastern regions presenting higher average values in walking distances. In total, only 0.1\% of the urban LSOAs fulfilled the three rules, with the North East having the highest number of regions that passed the rule.

\begin{figure}[H]
    \centering
    \includegraphics[width=\textwidth]{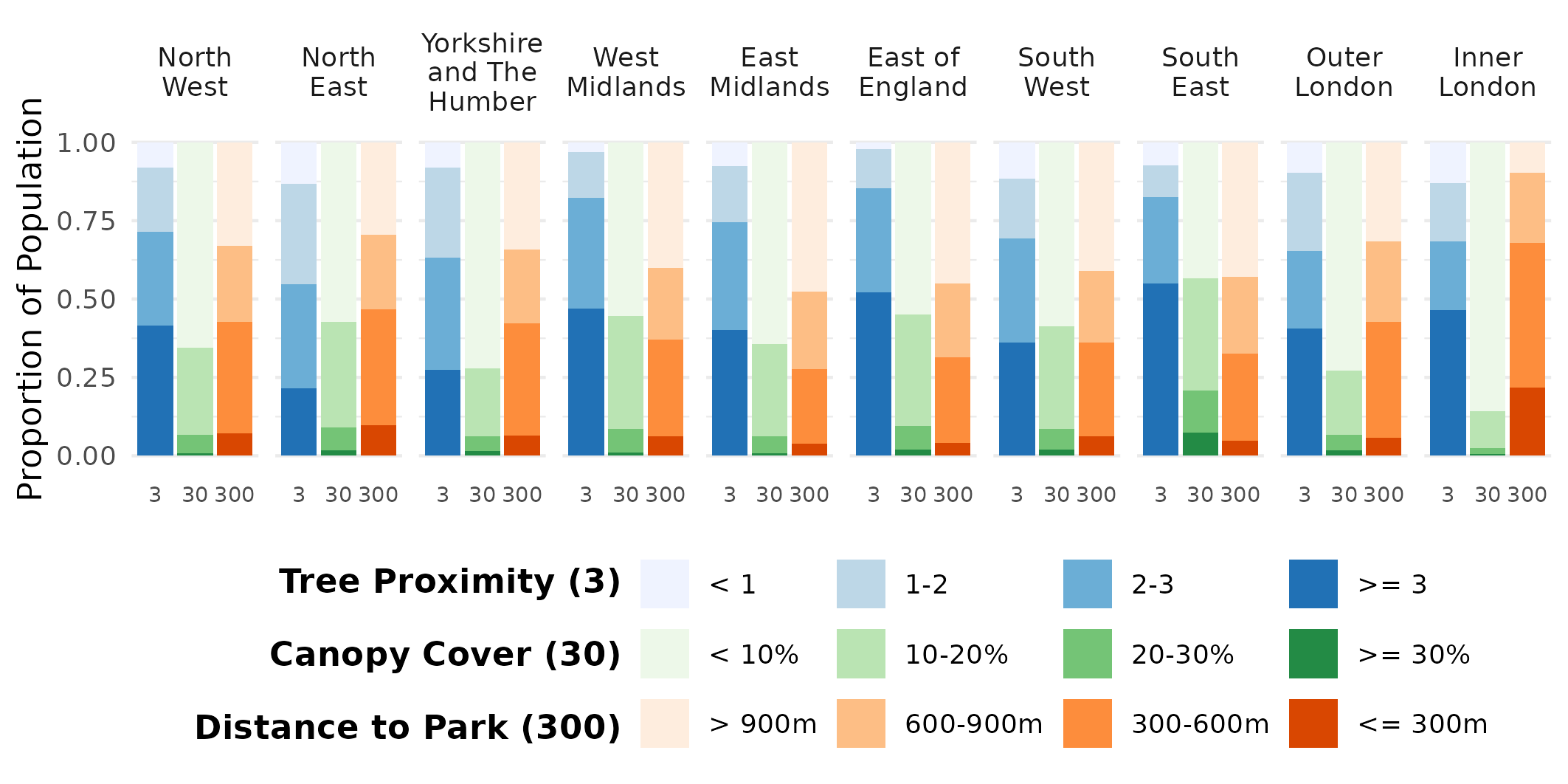}
    \caption{Population meeting each component of the 3-30-300 rule across the regions of England. The chart displays the proportion of the population in each region that fulfil each one of the rules (dark colours) and how close they are to achieving them (light colours): proximity to at least 3 trees in a 25 m-radius from a residence (3, blue), living in a neighbourhood with at least 30\% tree canopy cover (30, green), and living within 300 metres of a public park (300, orange).}
    \label{fig:3-30-300_population}
\end{figure}

Our national assessment reveals a profound disparity in the attainment rates for the three distinct components of the 3-30-300 rule across all regions of England (Figure \ref{fig:3-30-300_population}).

The ``3-tree proximity'' rule is the most widely achieved guideline. In London, for example, almost 40\% of the region's 8.2 million residents are estimated to live within 25 m of at least 3 trees from their homes, with no clear difference between inner and outer parts of the city. This pattern of high attainment holds true across the country, from the North West, where most of its 6.5 million people meet the standard, to the South West, indicating widespread success in integrating individual trees within highly urbanised areas. Nonetheless, approximately only one quarter of the inhabitants of Yorkshire and the Humber and the North East accomplish the goal.

The most significant national deficit is for the ``30\%-canopy cover'' rule, which is met by the smallest number of people. In every region, only a minority of the population lives in a neighbourhood with an adequate tree coverage, with the South East showing the largest proportion of urban inhabitants, followed by the North East, East of England, South West and Outer London.

The ``300 m- park accessibility'' rule demonstrates the second-highest level of attainment. While a substantial number of people meet this standard, the total is considerably less than for the 3-tree rule. London again has the largest number of residents where a public park is within walking distance, particularly in the central boroughs, whereas the East Midlands, East of England and South East show the largest number of inhabitants with low accessibility to green spaces.

\subsection{Systemic Environmental Inequality Gradients in Green Infrastructure}

\begin{figure}[H]
\centering
\includegraphics[width=\textwidth]{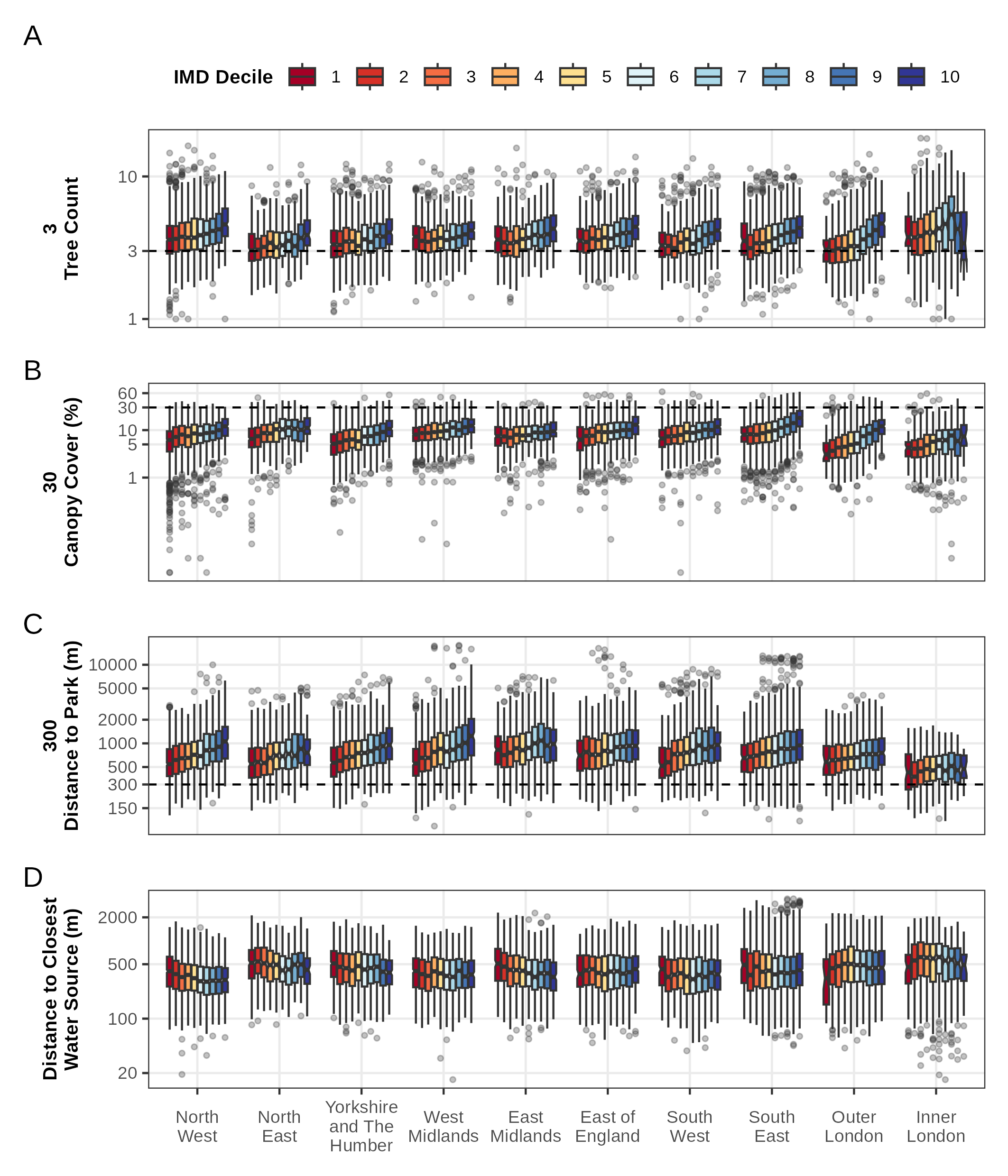}
\caption{Environmental inequality in access to green infrastructure across England. The distribution of the three 3-30-300 rule components at the Lower Layer Super Output Area (LSOA) level is shown, grouped by region. \textbf{(A)} Tree Count at 25 m radius. \textbf{(B)} Canopy cover percentage. \textbf{(C)} Walking distance to the nearest park. \textbf{(D)} Distance to the nearest water source. Dashed horizontal lines indicate the respective guideline thresholds for the 3-30-300. Each boxplot summarises the distribution for LSOAs within a given Index of Multiple Deprivation (IMD) decile, coloured from most deprived (Decile 1, red) to least deprived (Decile 10, blue); y-axes are log-scaled.}
\label{fig:3-30-300_imd}
\end{figure}

To dissect the relationship between socioeconomic status and green infrastructure, we analysed the distribution of each 3-30-300 component across Index of Multiple Deprivation (IMD) deciles for every region in England (Figure \ref{fig:3-30-300_imd}). The analysis reveals a clear, systematic, and contrasting pattern of environmental inequality for different types of green space.

For vegetation-based metrics, we observe a consistent green gap: canopy cover and tree proximity are strongly positively correlated with socioeconomic advantage. Within every single region, from the North West to London, LSOAs in the least deprived deciles have more trees in proximity (Figure \ref{fig:3-30-300_imd}A) and higher canopy cover (Figure \ref{fig:3-30-300_imd}B) than those in the most deprived deciles. For the London case, there is no clear difference in canopy cover, but there is more variability in proximity to trees.

In contrast, park accessibility exhibits a reverse green gap (Figure \ref{fig:3-30-300_imd}C). The most deprived LSOAs are, on average, located closer to a public park than the least deprived LSOAs. This trend is visible across most regions and suggests that the dense urban environments typically associated with higher deprivation have better proximity to public parks than more affluent suburban areas. Interestingly, inner boroughs in London have better access to parks. These findings expose a fundamental divergence in the distribution of urban green infrastructure: while formal public parks are highly accessible in dense, often deprived areas, ambient greenness like street trees and garden canopy is systematically skewed towards wealthier communities. 

In addition to the 3-30-300, water was also contrasted regionally (Figure \ref{fig:3-30-300_imd}D). Although it didn't show a strong correlation with deprivation, in some regions, such as the North West and the East Midlands, residents of affluent areas live closer to water sources, while in the rest of the regions, access to water is more uniform.

\subsection{The Geography of Greenness versus Inequality}

\begin{figure}[H]
\centering
\includegraphics[width=\textwidth]{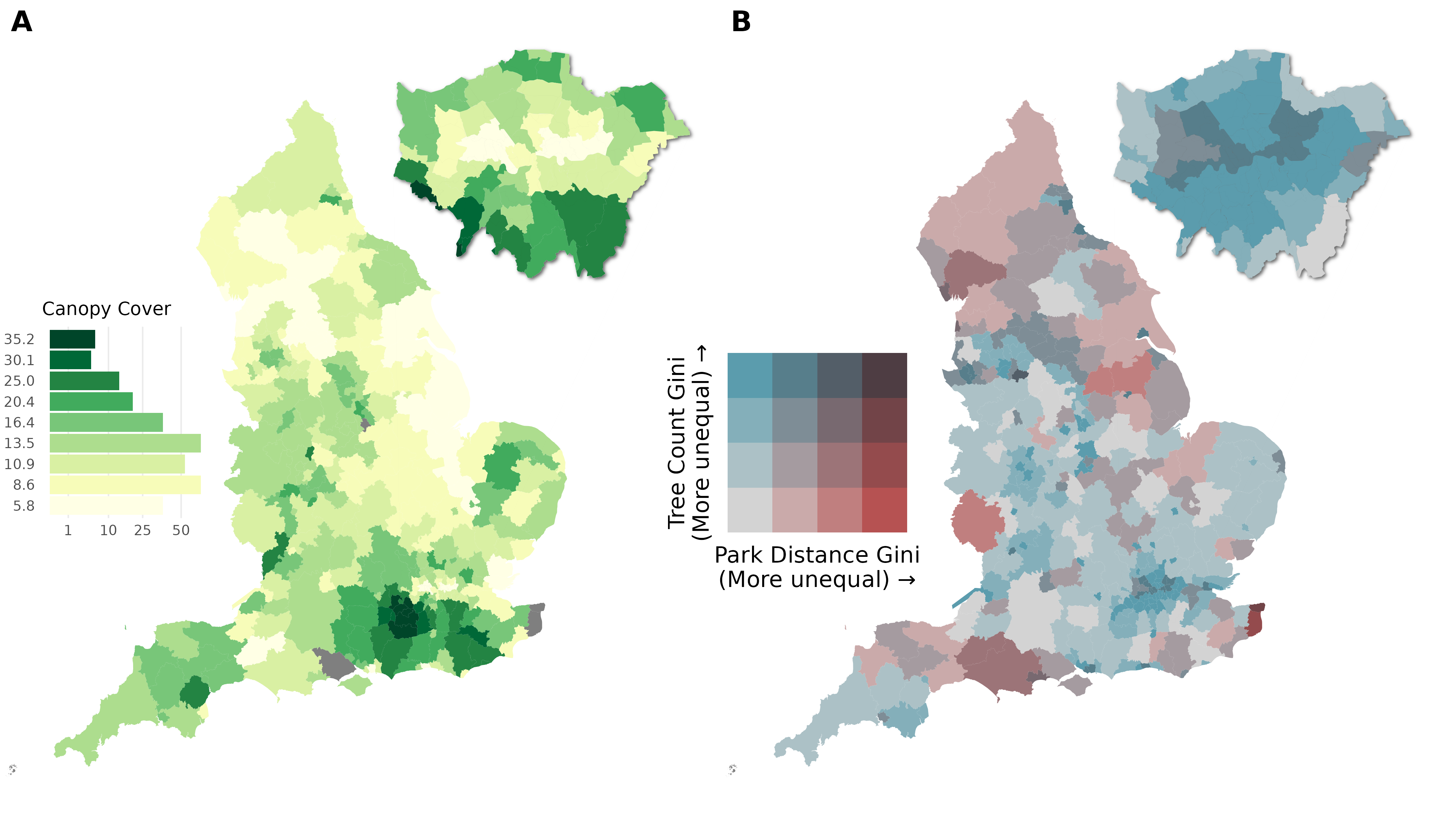}
\caption{Spatial Distribution of Canopy Cover and Environmental Inequality in England. \textbf{(A)} Average percentage of tree canopy cover at the Local Authority District (LAD) level. Darker green indicates higher canopy cover, with the highest values concentrated in the South East of England. The inset displays the significant variation across London's boroughs. \textbf{(B)} Bivariate map showing two measures of environmental inequality, calculated as a Gini coefficient at the LAD level. The colour scale indicates the degree of inequality in the distribution of nearby trees among buildings (blue y-axis) and inequality in walking distance to a park (red x-axis). Darker, mixed colours (e.g., purple) signify high inequality in both metrics.}
\label{fig:gini_maps}
\end{figure}

To understand the relationship between the quantity of green infrastructure and the equity of its distribution, we mapped both average canopy cover and the intra-regional inequality in access to nature across England (Figure \ref{fig:gini_maps}). The analysis of average canopy cover confirms that the South East of England is the nation's greenest region in absolute terms (Figure \ref{fig:gini_maps}A). Local Authorities in and around the London commuter belt, such as Surrey, consistently show the highest percentages of canopy cover, often approaching the 30\% guideline, while many districts in the Midlands and the North have lower average cover.

However, this picture of a green South East is fundamentally challenged by the geography of inequality (Figure \ref{fig:gini_maps}B). The bivariate analysis of Gini coefficients reveals that the most severe inequalities in green infrastructure access are often concentrated within these same leafy regions. Major urban centres—especially Inner London (see inset), Birmingham, and Manchester—emerge as dark hotspots, indicating high inequality in the distribution of both nearby trees (blue tones) and walking distance to parks (red tones) among their respective residents.

\subsection{Socioeconomic Stratification of Urban Environmental Metrics}

\begin{figure}[H]
\centering
\includegraphics[width=0.9\textwidth]{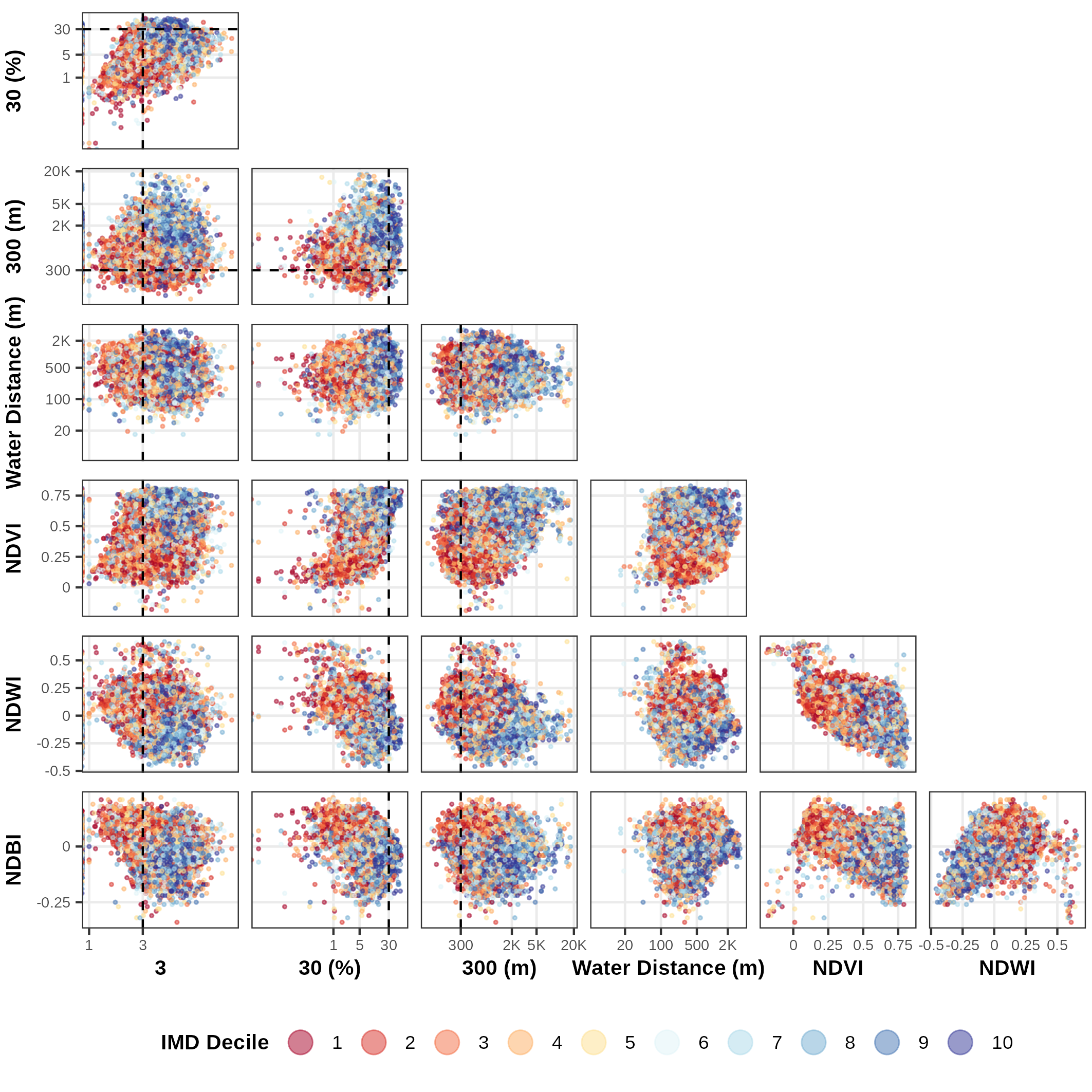}
\caption{Socioeconomic stratification of correlations between environmental metrics. The scatter plot matrix shows the pairwise relationships between the 3-30-300 rule components (3: Tree Count Index, 30: Canopy Cover, 300: Park Distance), distance to water, and satellite-derived spectral indices (NDVI, NDWI, NDBI). Each point is a Lower Layer Super Output Area (LSOA) in England, coloured by its Index of Multiple Deprivation (IMD) decile, from most deprived (Decile 1, red) to least deprived (Decile 10, blue). Dashed lines indicate the guideline thresholds for the 3-30-300 rule. Note that the axes for Park Distance and Water Distance are on a logarithmic scale.}
\label{fig:3-30-300_scatter_matrix}
\end{figure}

To understand how the relationships between different environmental metrics are shaped by socioeconomic status, we visualised their pairwise correlations, with each Lower Layer Super Output Area (LSOA) coloured by its Index of Multiple Deprivation (IMD) decile (Figure \ref{fig:3-30-300_scatter_matrix}). The analysis reveals a powerful, intertwined relationship between vegetation, built-up density, and deprivation. We observe a strong positive correlation between our vegetation-based rule components ('3' and '30') and satellite-derived Normalized Difference Built-up Index (NDVI) (Figure \ref{fig:ndvi_map}). Critically, these plots show a clear socioeconomic stratification: LSOAs with high NDVI and high canopy cover are almost exclusively the least deprived, while those with low vegetation and high Normalized Difference Built-up Index (NDBI) (Figure \ref{fig:ndbi_map}) scores are predominantly the most deprived.

The thresholds of the 3-30-300 rule further illustrate these disparities. A large number of the most deprived LSOAs fall below the 30\% canopy cover guideline, while a majority of the least deprived LSOAs meet or exceed it, confirming that failure to meet the vegetation-based components of the rule is systematically linked to socioeconomic deprivation, as demonstrated in the previous sections. This pattern is inverted when considering park accessibility. The plot of park distance ('300') versus NDBI shows that the most deprived LSOAs, which have the highest built-up density, are also clustered at shorter distances to parks. This corroborates the finding from the distributional analysis that park provision is often better in dense urban areas, even as ambient greenness is lower. Overall, the matrix demonstrates that environmental variables are not independent but form a nexus where deprivation is co-located with low ambient vegetation and high built-up density, despite often having good proximity to formal parks.

The analysis further reveals that proximity to blue space, as measured by water distance, represents a distinct dimension of environmental character with a low correlation to the green infrastructure metrics assessed. This suggests that the spatial distribution of blue infrastructure is driven by different factors than that of parks and tree canopy, and its relationship with deprivation is less direct. In contrast, the Normalized Difference Water Index (NDWI) (Figure \ref{fig:ndwi_map}) exhibited a positive correlation with NDVI, underscoring the frequent co-location of blue and green spaces within features like vegetated riparian zones. Finally, NDBI showed a strong negative correlation with both the '3' and '30' components. This powerful inverse relationship highlights how dense, grey urban infrastructure fundamentally constrains the available space for trees and canopy, quantitatively linking the most heavily built-up LSOAs with the lowest levels of ambient greenness.

In order to evaluate the relationship between remote sensing indices and the 3-30-300 variables, as well as water access, we fitted three spatial error models to identify the drivers of inequity for tree count, park distance, and water distance at the LSOA level. The significance of the spatial error term ($\lambda$) in all models ($p<2.2×10^{-16}$) confirmed strong spatial clustering (Table \ref{tab:sem_results}).

We found that more deprived areas, as measured by IMD suffer from greater inequity in tree distribution ($\beta=0.0004, p=0.003$). However, this effect is mainly a rural phenomenon. In urban areas, the relationship between deprivation and tree inequity is significantly weaker ($p=0.009$ for the interaction term). Inequity is also worse in highly built-up areas (NDBI) and places with less overall greenness (NDVI), while it's surprisingly lower in more densely populated LSOAs.

For park access, the trend reverses: higher IMD scores are linked to lower inequity ($\beta=-0.0005, p=0.010$). This suggests a pattern of uniformly poor access in more deprived areas rather than selective access. Urban areas generally have more equitable park access than rural ones. The strongest predictors were environmental, with inequity rising with population density and falling in greener areas (NDVI). The relationship was especially pronounced in London, where high deprivation was linked to even more equitable access.

The link between deprivation and access to water depends entirely on location. A powerful interaction ($p=1.3×10^{-9}$) shows two opposing trends: In rural areas, more deprivation is linked to lower inequity, while in urban areas, the relationship flips, and more deprivation is strongly associated with higher inequity. Overall, urban LSOAs have much higher baseline inequity in water access. This inequity is worsened by high population density and built-up land, and reduced by the presence of vegetation (NDVI) and water bodies (NDWI). Our findings highlight a critical urban-rural divide in environmental justice concerning blue spaces (Figure \ref{fig:gini_scatter}).

\section{Discussion} \label{sec:discussion}

In this study, we present the first national, standardised assessment of the 3-30-300 urban greening rule for England. Our analysis reveals a profound disconnect between different forms of nature access, challenging conventional narratives of environmental provision. Although England performs well on the “3” guideline—proximity to visible trees—it systematically fails to meet the “30” and “300” targets of neighbourhood canopy cover and walkable park access for most residents. These deficits are not evenly distributed: ambient greenness, such as tree canopy, is strongly associated with affluence, whereas proximity to parks is highest in denser, more deprived urban centres. The resulting pattern reflects a geography of environmental inequality, where the type and quality of nature one encounters are deeply stratified by wealth and settlement form.

This inequality unfolds across three intersecting gradients: regional, urban–rural, and socioeconomic. Southern regions, particularly the Southeast, exhibit higher tree density and canopy cover; however, this advantage is moderated by population density. Rural areas across both the north and south provide more trees per person, while urban cores concentrate people but not trees. Such patterns illustrate that the experience of greenness depends as much on demographic context as on geography itself. Within cities, socioeconomic divides amplify this variation. Affluent neighbourhoods enjoy both greater canopy cover and a more equitable distribution of trees, suggesting that investment in greening can deliver co-benefits for equity. In contrast, access to blue space exhibits the opposite trend—deprivation correlates strongly with both greater distance to and unequal access to water, demonstrating a persistent environmental disadvantage for low-income communities.

A particularly striking result is the inverse relationship between deprivation and park accessibility: the most deprived areas often comply most with the 300-metre rule. This counterintuitive finding underscores a proximity–quality paradox. High accessibility does not necessarily imply high-quality experiences; nearby parks in deprived areas may be smaller, less safe, or poorly maintained—factors that are often invisible to distance-based metrics. Thus, equating proximity with equitable access is an oversimplification. Future applications of the 3-30-300 rule should integrate qualitative dimensions—such as biodiversity, safety, and perceived amenity—to ensure that investments in green infrastructure translate into genuine social and health benefits.

The primary value of the 3-30-300 rule lies not in universal compliance but in its diagnostic capacity. Disentangling proximity, availability, and accessibility allows policymakers to pinpoint specific environmental deficits that composite indices like NDVI obscure. A city failing the 30 \% canopy target but passing the 300 m park threshold faces a different challenge than one in the reverse situation. Hence, rather than pursuing blanket attainment of all three targets, local authorities should combine these indicators with socioeconomic measures such as the Index of Multiple Deprivation (IMD) to design tailored, evidence-based greening strategies. This multidimensional approach provides a practical pathway to achieving SDG 11 by promoting equitable, context-sensitive urban sustainability.

Our methodological framework demonstrates the feasibility of conducting national-scale, building-level analyses using open LiDAR data and high-performance computing with Apache Sedona and Google Earth Engine. This creates a replicable blueprint for other nations seeking to assess environmental equity. Comparison with the Forest Research Trees Outside Woodland (TOW) dataset shows our approach identifies a higher number of tree features, owing to finer delineation of individual crowns rather than aggregated clusters. While this likely yields a more granular yet conservative estimate—London’s tree count, for instance, remains below independent estimates—such trade-offs are inherent to top-down approaches. Occlusion of smaller understory trees and the misclassification of shrubs remain key challenges. 

This study represents a foundational step toward more equitable urban greening; however, its true potential lies in its integration with emerging data sources. Future research should combine aerial and ground-level perspectives to capture the multi-dimensional experience of urban nature. High-resolution LiDAR and hyperspectral imagery can refine estimates of canopy and biodiversity, while street-level imagery and citizen-reported surveys can illuminate aspects of quality, safety, and usability that are invisible to remote sensing. Linking these layers within a common analytical framework would enable a more holistic assessment of urban ecosystems that accounts not only for the presence of green and blue infrastructure but also for how people perceive, use, and benefit from them. Embedding such data-rich approaches into national planning systems will be crucial for ensuring that future greening strategies deliver meaningful environmental and social value across all communities.

\section{Methods} \label{sec:methods}

Our study employs a multi-stage methodology to quantify the 3-30-300 rule at a national scale, assess its relationship with socioeconomic deprivation, and compare its components to traditional remote sensing indices. The workflow encompasses data acquisition, high-resolution geospatial processing, and statistical analysis of environmental inequality.

\subsection{Definition of Study Area} \label{ssec:study_area}

Most geographic statistics released by public agencies, particularly for census-related studies in Great Britain, are done using the  Lower Layer Super Output Areas (LSOAs) as a measurement unit, which is what we used in our study. However, only those that were in England according to the official December 2021 release by the Office for National Statistics (ONS) were considered.

\subsection{Socio-Economic and Demographic Indicators} \label{ssec:demography_data}

The Index of Multiple Deprivation (IMD) is a metric last produced in 2019 by the Ministry of Housing, Communities and Local Government for every LSOA in England. The index summarises seven main domains of inequality: Income, Employment, Education, Health, Crime, Housing and Environment. The scores for each component were downloaded from the Consumer Data Research Centre (CDRC) online platform. 

In addition, the mid-2022 edition of population estimates by the ONS, including age groups and gender counts, are included in the analysis. These variables are available at the LSOA level as well.

\subsection{3-30-300 Metrics} \label{ssec:3-30-300_data}

Most country-wide datasets for tree cover or individual tree locations are proprietary; therefore, our approach to measuring the '3-30-300' rule was based on using the publicly available 1-m-resolution Vegetation Object Model (VOM), gathered by the Environment Agency and published by  Department for Environment, Food \& Rural Affairs (Defra) as part of their Light Detection and Ranging (LiDAR) programme.

\subsubsection{Tree Segmentation} \label{sssec:tree_segmentation}

We first segmented individual trees on the VOM using the \textit{lidR} package in \textit{R} \cite{roussel_lidr_2020}. To segment the crown shape, the height was used with the Dalponte and \& Coomes algorithm \cite{dalponte_tree-centric_2016} and a customised formula for the Local Maximum Filter (LMF), as seen in Equation \ref{eq:lidar_formula}, where $z$ is the height in metres for a given pixel, and the result is the window size to check for minimum and maximum values to model the crown shape. This approach is more adaptive than a fixed-size filter, as it enables the algorithm to search for smaller crowns in areas with dense, young trees and larger crowns in mature, isolated trees, thereby better reflecting the real-world forest structure. While methods with higher geometric fidelity exist for single-tree analysis, our approach is optimised for large-scale quantification. Its use of a simplified crown model, combined with the dataset's broad spatial range, offers a consistent methodology for quantifying trees without introducing systematic bias to the overall findings. 

\begin{equation} \label{eq:lidar_formula}
\text{LMF}(z) =
\begin{cases} 
\text{min\_size}, & \text{if } s < \text{min\_size} \\
s, & \text{if } \text{min\_size} \leq s \leq \text{max\_size} \\
\text{max\_size}, & \text{if } s > \text{max\_size}
\end{cases}
\end{equation}

Where:

\[
s = \left\lfloor 6 + 18 \cdot \exp\left(-\frac{(z - \mu)^2}{2\sigma^2}\right) \right\rfloor
\]

With:  

- \( z \) = canopy height at a given location 

- \( \mu = 18 \) (mean of the Gaussian function)  

- \( \sigma = 7 \) (spread of the Gaussian function)  

- \( \text{min\_size} = 7 \) (minimum search window size)  

- \( \text{max\_size} = 0.7 \times P_{95}(z) \) (maximum search window based on the 95th percentile)\\

The resulting trees were vectorised as points and grouped together into the 50x50-km tiles, following the definition in the British National Grid released by the Ordnance Survey (OS), which were then saved as geoparquet files.

\subsubsection{3 Component} \label{sssec:3_data}

We define the first component of the rule (3) as tree proximity. We created buffers of different sizes (10, 25, 50, 75 and 100 m) around each feature in the Verisk buildings dataset and counted the number of trees inside that area. This dataset includes features such as height, number of floors and distance to water, which this study uses to quantify water access. Due to the number of polygons in the buildings layer, this step was performed using \textit{Apache Sedona} RDD API in \textit{Python} alongside a vectorised and point-based version of the VOM-derived tree product. This spatial join operation was performed using QuadTree spatial indexing and KDBTree partitioning to optimise computation times.

\subsubsection{30 Component} \label{sssec:30_data}

Canopy cover or green space availability (30) was obtained by creating a binary layer of the VOM raster where pixels between 3 and 60 m were considered 1. Then, using the LSOA boundaries, we calculated the tree coverage at 1 m resolution. 

\subsubsection{300 Component} \label{sssec:300_data}

Finally, for park accessibility (300), we filtered the OS Green Space Layer to include only public parks and calculated the walking distance (network-based) from each building to the closest green space access point. To accomplish this, a road network was built from the OS Roads dataset, and Dijkstra's algorithm was used to measure the shortest path along the graph between each building and its corresponding park.

Due to the size of the datasets, particularly of the buildings and VOM-derived trees, the results of the three components were calculated using Local Authority District (LAD) geometries, where the datasets were clipped using the LAD to reduce the number of iterations. Finally, the 3 and 300 components, measured at the building level, were aggregated at the LSAO level to match the IMD spatial unit of measurement as that of the 30-component. In addition to this, a total count of trees was done at the same geographic level.

\subsection{Spectral Indices} \label{ssec:spectral_data}

Three main (normalised) spectral indices were considered in this study. These indicators were calculated using Sentinel 2 scenes from 2024 with values under 10\% cloud coverage through the Google Earth Engine \textit{Python} API \cite{montero_standardized_2023}. These were estimated as the maximum for the entire year for the study region. To aggregate the values at the LSOA level, we used the median value for all pixels falling into the polygon.\\

%\begin{table}[h!]
%\caption{Normalised spectral indices calculated using Sentinel 2 scenes.}
%\label{tab:spectral_index}

\begin{tabular}{llll}
\hline
Index & Name & Formula & Reference \\ \hline
NDBI & Normalized Difference Built-Up Index & $\frac{B11-B8}{B11+B8}$ & \cite{zha_use_2003, zheng_improved_2021} \\
NDVI & Normalized Difference Vegetation Index & $\frac{B8-B4}{B8+B4}$ & \cite{martinez_demystifying_2023, li_comparison_2015} \\
NDWI & Normalized Difference Water Index & $\frac{B3-B8}{B3+B8}$ & \cite{yang_mapping_2017, zheng_improved_2021}
\end{tabular}
%\end{table}

\subsection{Data Synthesis and Scale of Analysis} \label{subsec:data_synthesis}
To conduct the first national-scale assessment of the 3-30-300 rule, we gathered multiple high-resolution geospatial datasets. This process created a comprehensive building-level record of potential nature exposure for the entirety of England, representing an analysis of unprecedented scale and granularity. The core components of this synthesised dataset are summarised in Table \ref{tab:data_synthesis}, quantifying the millions of individual features processed to derive our findings.

\begin{table}[h!]
\centering
\caption{Data Synthesis and Scale of the National 3-30-300 Assessment. This table outlines the primary datasets, unit of analysis, and overall scale for each analytical component quantified across England.}
\label{tab:data_synthesis}
% Using standard 'tabular' with 'p' columns for manual text wrapping
\begin{tabular}{p{2.7cm} p{3cm} p{2.2cm} p{3.5cm}} 
\toprule
\textbf{Analytical Component} & \textbf{Core Datasets Used} & \textbf{Initial Unit of Measurement} & \textbf{Scale of Analysis (No. of Features)} \\
\midrule
\textbf{3} (Proximity) & VOM, Buildings & Individual Building & \textasciitilde190 million trees; \newline 28,944,175 buildings \\
\addlinespace 

\textbf{30} (Availability) & VOM, LSOA Boundaries & 1m\textsuperscript{2} Raster Grid Cell & 33,755 LSOAs \\
\addlinespace

\textbf{300} (Accessibility) & OS Roads, Green Spaces, Verisk Buildings & Individual Building & 3,919,444 road segments; \newline 157,274 green spaces \newline 28,944,175 buildings \\
\addlinespace

\textbf{Socio-Economic Context} & Deprivation Score & LSOA Polygon & 32,742 LSOAs \\
\addlinespace

\textbf{Spectral Indices} & NDVI, NDWI, NDBI & 100m\textsuperscript{2} Raster Grid Cell & Full coverage of England \\
\bottomrule
\end{tabular}
\end{table}

\subsection{Inequality Measurement}

The 3 and 300 components measured in the study, as well as the distance to water variable in the Verisk buildings dataset, were used to calculate the Gini Index at the LSOA level. To do so, the unaggregated measurements for each building categorised as a residential unit were considered in the analysis. The Gini coefficient was calculated using the DescTools package in R using the unbiased parameter, which corrects for differences in sample size, as shown in \ref{eq:gini_formula}. This correction makes the coefficient to have a range between 0 and $\sqrt{\frac{n-1}{n}}$, unlike regular Gini metrics that can only be between 0 and 1.

\begin{equation} \label{eq:gini_formula}
G_{\text{unbiased}} = \left( \frac{\sum_{i=1}^{n} \sum_{j=1}^{n} |x_i - x_j|}{2n \sum_{i=1}^{n} x_i} \right) \times \frac{n}{n-1}
\end{equation}

With:

- \( n \) = number of buildings in a given LSOA

- \( x_i \) and \( x_j \) = environmental metric for two buildings\\

To create a single, robust metric for the ``3-tree proximity'' component, we moved beyond a simple count at a single distance. For each building, we counted the VOM-derived trees within five concentric buffers (10, 25, 50, 75, 100 m). We then fitted an exponential regression to these counts against the buffer radius (Figure \ref{fig:tree_slope_plots}). The slope of this regression was used as our final metric. This approach offers a more nuanced measure of tree availability than a simple count; a steeper slope indicates not just the presence of trees, but a rapid increase in their number as one's view expands outwards, proxying for a richer, denser tree-dominated area.

The canopy cover component was not considered in the inequality estimations because it was not measured at the building level unlike the other metrics. Because the Gini coefficient is a measure of inequality for one area (e.g. LSOA), it requires a more granular sampling unit (e.g. residential building) from which to derive the cumulative (green) wealth.

\subsection{Statistical Modelling} \label{ssec:statistical_modelling}
To analyse the socio-demographic drivers of environmental inequality, we addressed the inherent spatial autocorrelation in our LSOA-level data using a Spatial Error Model (SEM), which models spatial dependence in the error term, to account for unobserved, spatially structured covariates. Separate models were fitted for each of our three inequality outcome variables: the Gini coefficients for tree distribution (slope of the exponential regression), park access (walking distance to closest park), and blue space access (straight-line distance to closest water source).

The predictor variables for each model included population density, urban-rural classification, region, the Index of Multiple Deprivation (IMD) score, and mean spectral indices (NDVI, NDWI, NDBI). To investigate how the relationship between deprivation and green/blue space inequality varies across different contexts, we incorporated interaction terms between the IMD score and both urban-rural classification and region. The generalised model is formalised in Equation \ref{eq:sem_equation}.

\begin{equation} \label{eq:sem_equation}
\begin{split}
G_{unbiased} &\sim \text{PopDensity} + \text{NDVI} + \text{NDWI} + \text{NDBI} +\\
&\text{IMD} * \text{UrbanRural} + \text{IMD} * \text{Region}
\end{split}
\end{equation}

\subsection{Limitations} \label{ssec:limitations}
While our methodology is capable of measuring each individual component of the 3-30-300 for all the buildings, it does so using a general assumption of tree shapes from the VOM, which creates square-shaped polygons that don't represent the real shape of trees. Moreover, this is calculated from a mosaic of LiDAR reads collected from 2018 to 2023; hence, changes in urban canopy due to tree removal might not be accounted for. 

Measurement of tree visibility requires view-shed modelling, which involves a 3D representation of each building and its surroundings. This method is unfeasible for the number of buildings and trees used, thus why we defined the 3 metric as tree proximity instead of visibility.

Canopy cover was estimated using the VOM, which represented the highest-resolution and openly available vegetation product at the national level at the time of this study. However, it is worth noting that privately owned datasets of trees that include crown area and height exist, which could be used as a more accurate representation of urban trees. 

In addition, some tiles from the original source were corrupted or missing, which limited the calculation for certain areas, affecting both 3- and 30-component models that relied on the VOM product (Figure \ref{fig:data_coverage_map}). 

In a similar fashion, the 300 calculation was impacted by the spatial extent of the processing, meaning that for a given LAD with no public green spaces, the estimate for all buildings would be null, or if there is no direct path from a building to a park in the road network, the aggregated value would not be determined. Vehicle roads were used to measure the distance to the closest access point to an open public park; however, walking paths might differ in length from those used by vehicles. 

Finally, although the Verisk dataset we used is proprietary, an open version with the same building polygons is available through the EDINA platform, which academic institution members can access fully for research purposes.

%%A graphical summary of the methodology is presented in Figure \ref{fig:t3_30_300_methods}

%%\begin{figure}[H]
%%\centering
%%\includegraphics[width=\textwidth]{t3_30_300_methods.png}
%%\caption{Overview of the data integration and processing workflow. The methodology synthesises multiple national datasets (right), including the Vegetation Object Model (VOM) and Ordnance Survey (OS) layers for roads, buildings, and green spaces. These inputs are used to calculate the three components of the 3-30-300 rule (left): park accessibility (300m) via network analysis, canopy cover availability (30\%) from the binarised VOM, and visible tree counts (3) from LiDAR-based tree segmentation.}
%%\label{fig:t3_30_300_methods}
%%\end{figure}

\section{Code and Data Availability} \label{sec:data}
The Python and R code used to measure the 3-30-300 rule and extract data from Sentinel 2 images is available in the GitHub repository (\href{https://github.com/ancazugo/3-30-300-analysis}{https://github.com/ancazugo/3-30-300-analysis}). The aggregated data at LSOA and LAD levels are available in the Zenodo repository (\href{https://doi.org/10.5281/zenodo.16911970}{https://doi.org/10.5281/zenodo.16911970}). Tree segmentation tiles are available upon request. Two interactive visualisation demos are available on the \href{https://ancazugo.github.io/3-30-300-analysis/}{website} for the code repository: the first one displays aggregated data at the LSOA level, and the second one shows the location of all trees in England.

\begin{appendices}

\section{Supplementary Material}

\begin{figure}[H]
    \centering
    \includegraphics[width=1\linewidth]{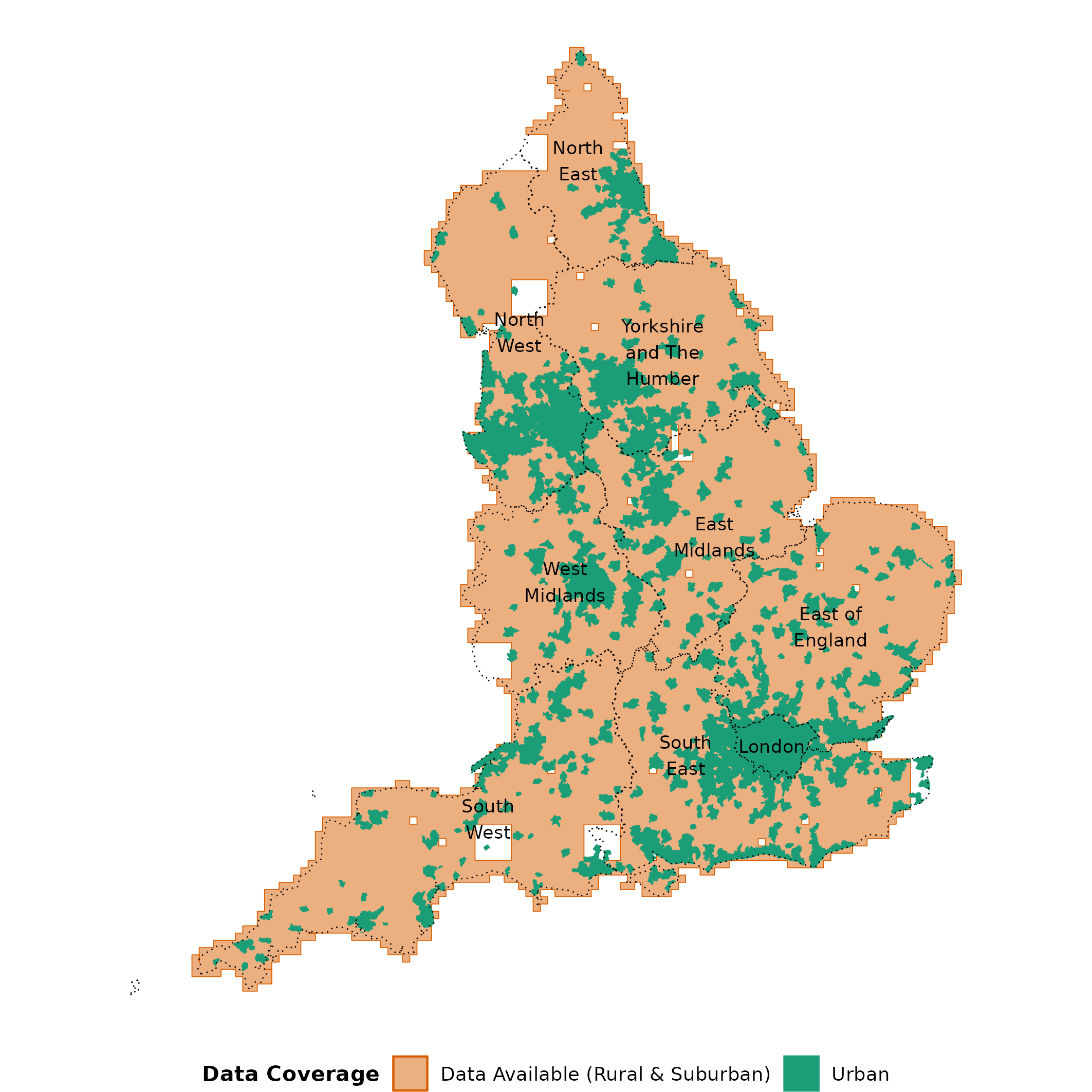}
    \caption{Coverage of the tree segmentation process in all of England, including rural and urban areas. Blank patches represent areas where data was missing or corrupted.}
    \label{fig:data_coverage_map}
\end{figure}

\begin{table}
    \centering
    \caption{Summary of the number of urban Lower Layer Super Output Areas (LSOAs) per region that fulfil each component of the 3-30-300 rule and all of them combined. The values for the components (3, 30, 300) are percentages of the total urban LSOAs in that region, while the 3-30-300 column represents the percentage of LSOAs fulfilling all criteria simultaneously.}
    \sisetup{
        table-format=2.1, % Sets the number format for the S columns
    }
    \begin{tabular}{
        l           % Left-aligned column for Region names
        S[table-format=5.0] % Integer column for # Urban LSOAs
        S           % S column for '3'
        S           % S column for '30'
        S           % S column for '300'
        S[table-format=1.1] % S column for '3-30-300'
    }
        \toprule
        % Column Headers
        {Region} & {No. Urban LSOAs} & {3 (\%)} & {30 (\%)} & {300 (\%)} & {3-30-300 (\%)} \\
        \midrule
        North West & 4135 & 41.0 & 0.7 & 7.4 & 0.0\\
North East & 1393 & 21.4 & 1.7 & 10.1 & 0.1\\
Yorkshire and The Humber & 2802 & 27.5 & 1.4 & 6.7 & 0.0\\
West Midlands & 3062 & 45.7 & 1.3 & 7.1 & 0.1\\
East Midlands & 2122 & 39.6 & 0.8 & 4.3 & 0.0\\
East of England & 2740 & 51.5 & 2.1 & 4.4 & 0.0\\
South West & 2411 & 35.8 & 2.0 & 6.4 & 0.2\\
South East & 4542 & 53.9 & 7.4 & 5.4 & 0.1\\
London & 4994 & 43.3 & 1.3 & 12.7 & 0.0\\
        \midrule
        % The last row is set in bold face
        \bfseries England        & \bfseries 28201 & \bfseries 42.2 & \bfseries 2.3 & \bfseries 7.4 & \bfseries 0.1 \\
        \bottomrule
        \label{tab:3-30-300_table}
    \end{tabular}
\end{table}

\begin{figure}[H]
    \centering
    \includegraphics[width=1\linewidth]{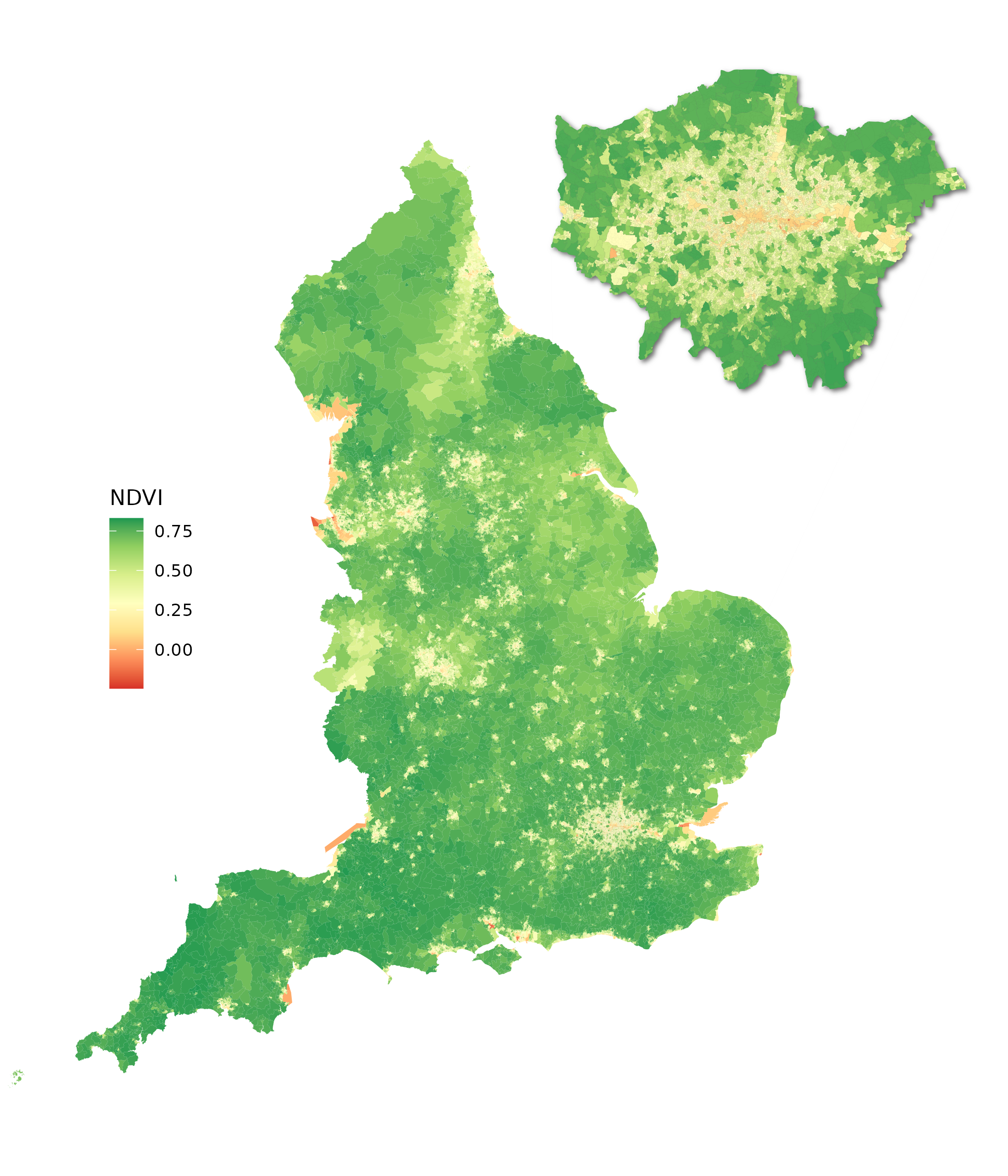}
    \caption{NDVI for each LSOA in England}
    \label{fig:ndvi_map}
\end{figure}

\begin{figure}[H]
    \centering
    \includegraphics[width=1\linewidth]{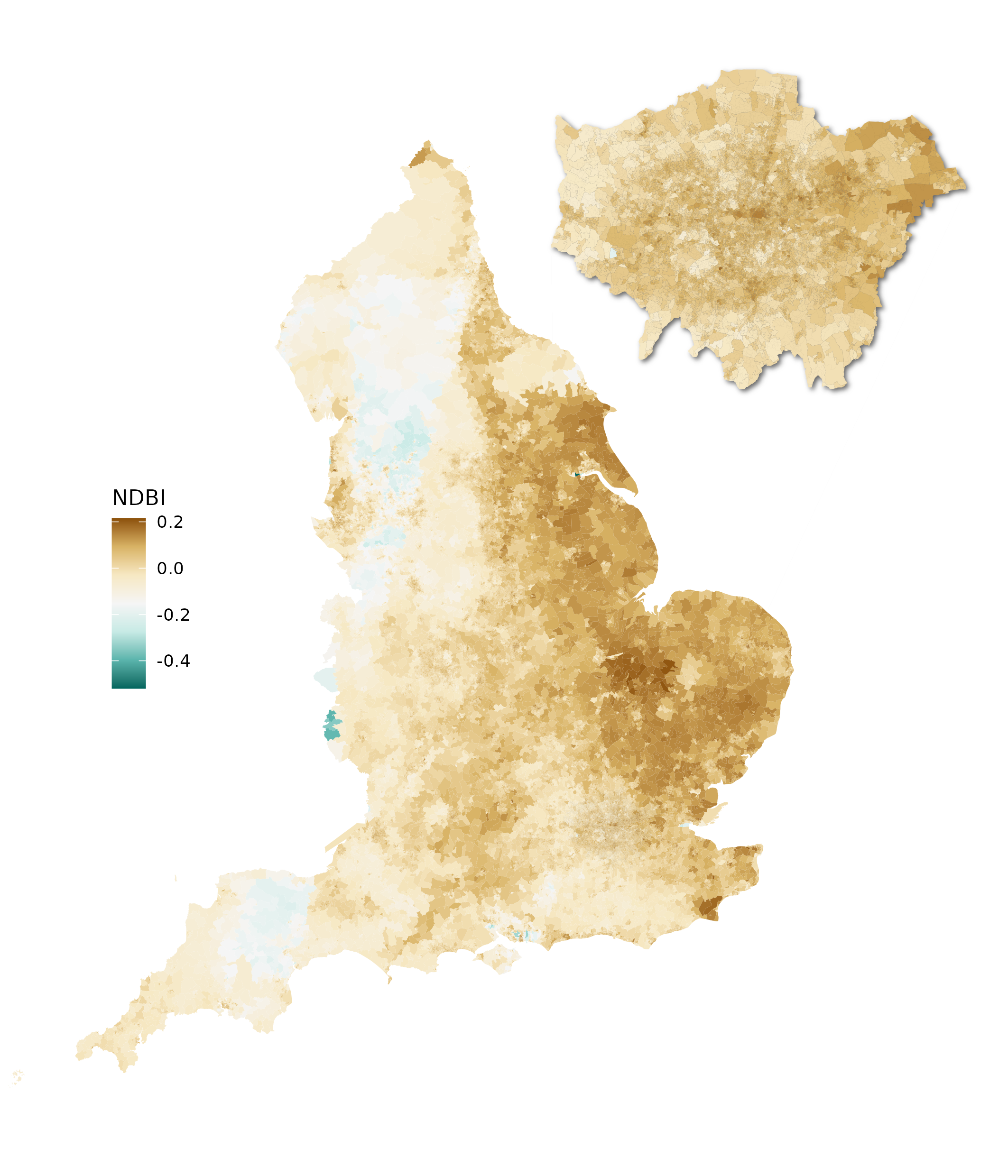}
    \caption{NDBI for each LSOA in England}
    \label{fig:ndbi_map}
\end{figure}

\begin{figure}[H]
    \centering
    \includegraphics[width=1\linewidth]{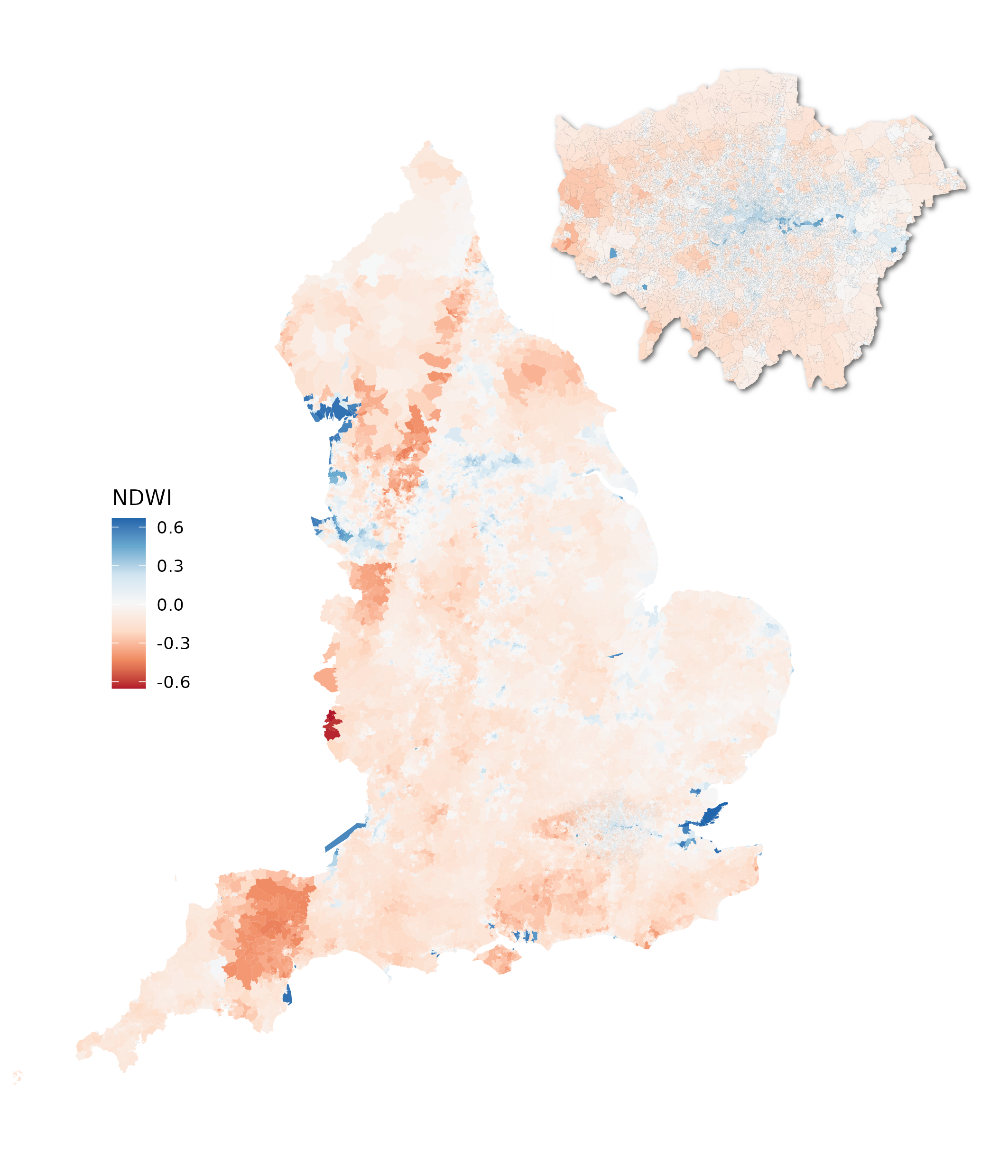}
    \caption{NDWI for each LSOA in England}
    \label{fig:ndwi_map}
\end{figure}

\begin{landscape}
\begin{table}[htbp]
\centering
\caption{Spatial error model results for inequality (Gini coefficients) in tree count, park distance, and water distance at the LSOA level in England. Standard errors in parentheses. Significance: * $p<0.05$, ** $p<0.01$, *** $p<0.001$.}
\label{tab:sem_results}
\begin{tabular}{lcccccc}
\hline
 & \multicolumn{2}{c}{Tree Count Gini} & \multicolumn{2}{c}{Park Distance Gini} & \multicolumn{2}{c}{Water Distance Gini} \\
Variable & Estimate & SE & Estimate & SE & Estimate & SE \\
\hline
(Intercept) & 0.13293*** & (0.00464) & 0.84046*** & (0.00622) & 0.72883*** & (0.00607) \\
IMDScore & 0.00041** & (0.00014) & -0.00052* & (0.00020) & -0.00053** & (0.00019) \\
UrbanUrban & -0.00168 & (0.00245) & -0.01647*** & (0.00350) & 0.04937*** & (0.00333) \\
RegionNorth East & -0.00389 & (0.00604) & -0.02246** & (0.00717) & 0.03714*** & (0.00739) \\
RegionYorkshire and The Humber & -0.00694 & (0.00483) & 0.00046 & (0.00576) & 0.01955** & (0.00594) \\
RegionWest Midlands & -0.03296*** & (0.00478) & 0.01479** & (0.00570) & 0.00527 & (0.00588) \\
RegionEast Midlands & -0.02556*** & (0.00500) & 0.01338* & (0.00601) & 0.00995 & (0.00617) \\
RegionEast of England & -0.04834*** & (0.00478) & 0.00405 & (0.00577) & 0.01542** & (0.00591) \\
RegionSouth West & -0.00706 & (0.00486) & 0.00918 & (0.00577) & -0.00241 & (0.00596) \\
RegionSouth East & -0.04572*** & (0.00425) & 0.00627 & (0.00505) & 0.01465** & (0.00521) \\
RegionLondon & -0.04834*** & (0.00475) & 0.00471 & (0.00581) & 0.02851*** & (0.00592) \\
population\_density & -0.00000** & (0.00000) & 0.00000*** & (0.00000) & 0.00001*** & (0.00000) \\
NDVI & -0.03152*** & (0.00450) & -0.06172*** & (0.00642) & -0.06950*** & (0.00612) \\
NDWI & 0.02578*** & (0.00577) & -0.01505 & (0.00794) & -0.06503*** & (0.00771) \\
NDBI & 0.14228*** & (0.01086) & 0.00453 & (0.01504) & 0.05381*** & (0.01458) \\
IMDScore:UrbanUrban & -0.00032** & (0.00012) & 0.00023 & (0.00018) & 0.00103*** & (0.00017) \\
IMDScore:RegionNorth East & 0.00011 & (0.00013) & 0.00029 & (0.00018) & 0.00002 & (0.00017) \\
IMDScore:RegionYorkshire and The Humber & -0.00010 & (0.00011) & -0.00010 & (0.00015) & 0.00010 & (0.00015) \\
IMDScore:RegionWest Midlands & -0.00035** & (0.00011) & -0.00035* & (0.00016) & -0.00029 & (0.00015) \\
IMDScore:RegionEast Midlands & -0.00025 & (0.00013) & 0.00026 & (0.00018) & 0.00011 & (0.00017) \\
IMDScore:RegionEast of England & -0.00000 & (0.00013) & 0.00013 & (0.00019) & -0.00034 & (0.00018) \\
IMDScore:RegionSouth West & -0.00030* & (0.00013) & -0.00021 & (0.00019) & -0.00029 & (0.00018) \\
IMDScore:RegionSouth East & 0.00002 & (0.00012) & -0.00005 & (0.00017) & -0.00035* & (0.00017) \\
IMDScore:RegionLondon & 0.00011 & (0.00014) & -0.00079*** & (0.00019) & -0.00089*** & (0.00018) \\
\hline
\multicolumn{7}{l}{\textbf{Model diagnostics}} \\
$\lambda$ & 0.64497 &  & 0.45495 &  & 0.55726 &  \\
Asymptotic SE ($\lambda$) & 0.00515 &  & 0.00652 &  & 0.00583 &  \\
Wald statistic ($\lambda$) & 15701.00 &  & 4863.30 &  & 9120.60 &  \\
Log likelihood & 42760.54 &  & 29928.67 &  & 32469.97 &  \\
ML residual variance & 0.00343 &  & 0.00827 &  & 0.00683 &  \\
AIC & -85469.00 &  & -59805.00 &  & -64888.00 &  \\
\hline
\end{tabular}
\end{table}
\end{landscape}

\begin{figure}[htbp]
    \centering
    \includegraphics[width=.9\linewidth]{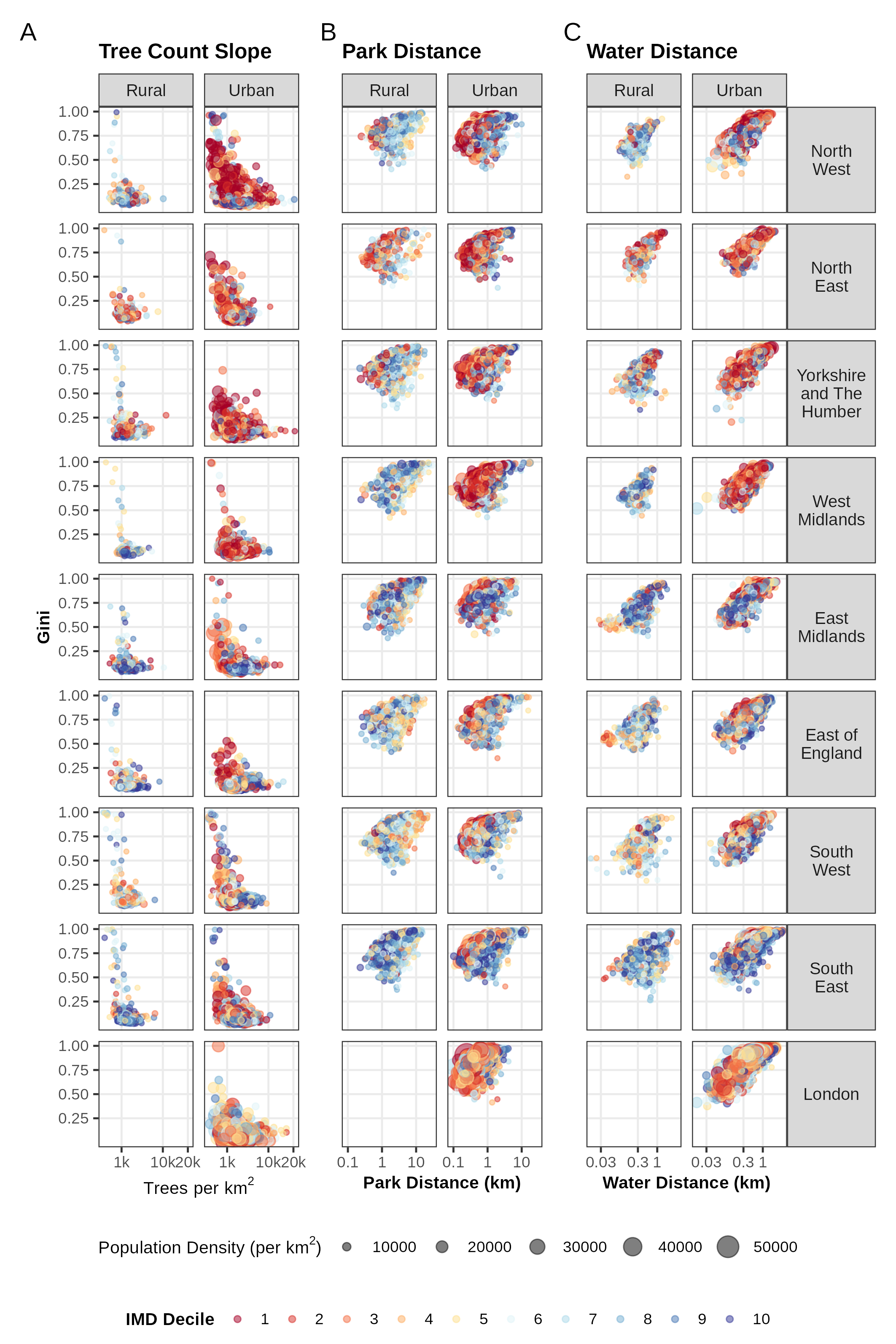}
    \caption{Scatter plot between the three Gini coefficients and their corresponding environmental metric for each region in England. Size depicts population density, while colours point to IMD classification. A: Tree count slope, B: Walking distance to closest park, and C: Distance to closest water source.}
    \label{fig:gini_scatter}
\end{figure}

\begin{table}[h!]
    \centering
    \caption{Comparison of tree geometry counts and total area between the Forest Research Trees Outside Woodland (TOW) dataset and this study's segmentation approach, by region.}
    \label{tab:tow_comparison_detailed}
    \begin{tabular}{lrrrr}
    \toprule
    \multirow{2}{*}{\textbf{Region}} & \multicolumn{2}{c}{\textbf{TOW Dataset}} & \multicolumn{2}{c}{\textbf{This Study}} \\ 
    \cmidrule(lr){2-3} \cmidrule(lr){4-5}
    & \textbf{Count} & \textbf{Total Area (m²)} & \textbf{Count} & \textbf{Total Area (m²)} \\
    \midrule
    North West & 4,155,143 & 613,505,095 & 8,205,613 & 1,005,555,299 \\
    North East & 1,571,966 & 229,361,711 & 2,263,592 & 699,848,693 \\
    Yorkshire and the Humber & 3,444,172 & 534,527,559 & 6,425,401 & 1,040,430,826 \\
    West Midlands & 4,252,559 & 833,297,949 & 5,164,030 & 1,223,817,794 \\
    East Midlands & 3,778,562 & 653,799,830 & 4,897,732 & 1,060,999,126 \\
    East of England & 4,941,670 & 959,028,854 & 6,166,459 & 1,718,896,157 \\
    South West & 6,583,686 & 1,474,924,354 & 4,911,044 & 2,512,667,899 \\
    South East & 6,157,704 & 1,370,145,617 & 9,419,679 & 2,718,374,849 \\
    London & 1,418,131 & 167,073,513 & 3,689,884 & 128,330,021 \\
    \bottomrule
    \end{tabular}
\end{table}

\begin{figure}[H]
\centering
\includegraphics[width=\textwidth]{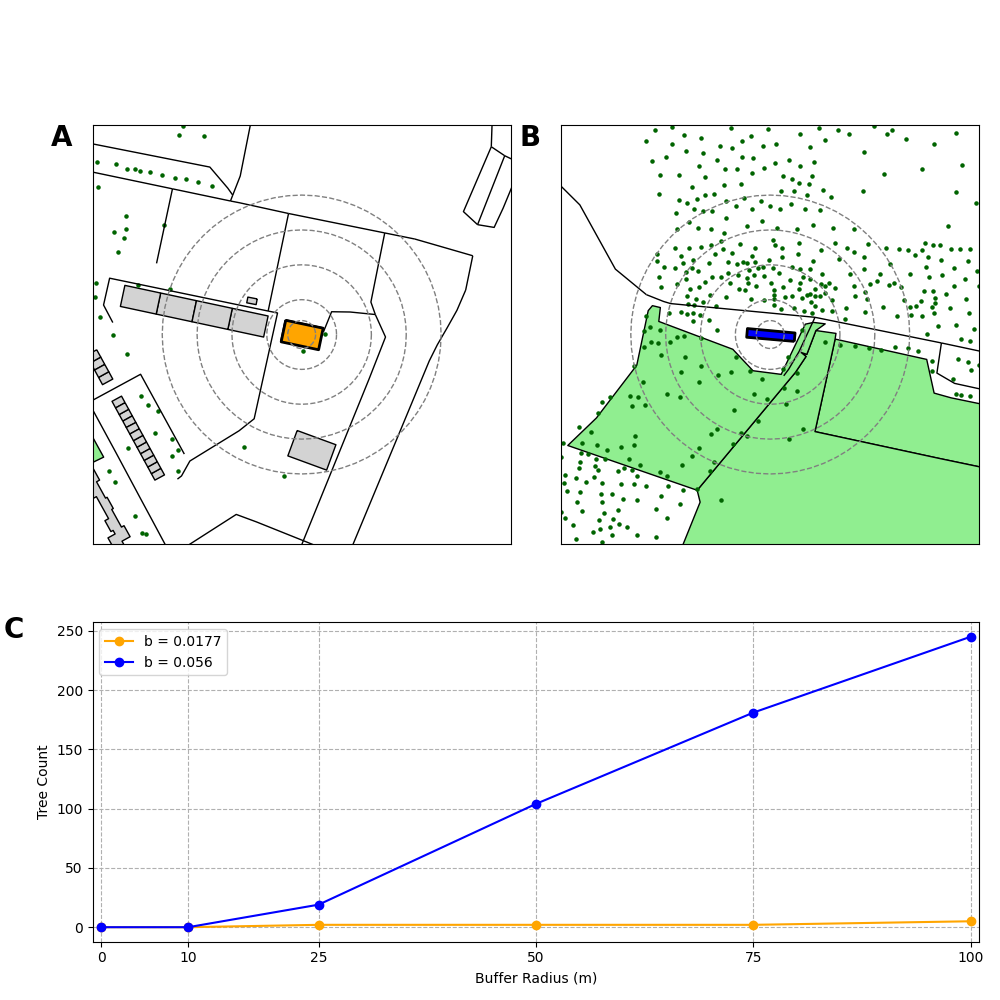}
\caption{Calculation of the tree index using the buffered count. A depicts a residential unit with a low slope in the regression, while B represents a house with a high slope. C shows the regression for both buildings.}
\label{fig:tree_slope_plots}
\end{figure}

\end{appendices}


\bibliography{references}% common bib file


\begin{thebibliography}{26}
\ifx \bisbn   \undefined \def \bisbn  #1{ISBN #1}\fi
\ifx \binits  \undefined \def \binits#1{#1}\fi
\ifx \bauthor  \undefined \def \bauthor#1{#1}\fi
\ifx \batitle  \undefined \def \batitle#1{#1}\fi
\ifx \bjtitle  \undefined \def \bjtitle#1{#1}\fi
\ifx \bvolume  \undefined \def \bvolume#1{\textbf{#1}}\fi
\ifx \byear  \undefined \def \byear#1{#1}\fi
\ifx \bissue  \undefined \def \bissue#1{#1}\fi
\ifx \bfpage  \undefined \def \bfpage#1{#1}\fi
\ifx \blpage  \undefined \def \blpage #1{#1}\fi
\ifx \burl  \undefined \def \burl#1{\textsf{#1}}\fi
\ifx \doiurl  \undefined \def \doiurl#1{\url{https://doi.org/#1}}\fi
\ifx \betal  \undefined \def \betal{\textit{et al.}}\fi
\ifx \binstitute  \undefined \def \binstitute#1{#1}\fi
\ifx \binstitutionaled  \undefined \def \binstitutionaled#1{#1}\fi
\ifx \bctitle  \undefined \def \bctitle#1{#1}\fi
\ifx \beditor  \undefined \def \beditor#1{#1}\fi
\ifx \bpublisher  \undefined \def \bpublisher#1{#1}\fi
\ifx \bbtitle  \undefined \def \bbtitle#1{#1}\fi
\ifx \bedition  \undefined \def \bedition#1{#1}\fi
\ifx \bseriesno  \undefined \def \bseriesno#1{#1}\fi
\ifx \blocation  \undefined \def \blocation#1{#1}\fi
\ifx \bsertitle  \undefined \def \bsertitle#1{#1}\fi
\ifx \bsnm \undefined \def \bsnm#1{#1}\fi
\ifx \bsuffix \undefined \def \bsuffix#1{#1}\fi
\ifx \bparticle \undefined \def \bparticle#1{#1}\fi
\ifx \barticle \undefined \def \barticle#1{#1}\fi
\bibcommenthead
\ifx \bconfdate \undefined \def \bconfdate #1{#1}\fi
\ifx \botherref \undefined \def \botherref #1{#1}\fi
\ifx \url \undefined \def \url#1{\textsf{#1}}\fi
\ifx \bchapter \undefined \def \bchapter#1{#1}\fi
\ifx \bbook \undefined \def \bbook#1{#1}\fi
\ifx \bcomment \undefined \def \bcomment#1{#1}\fi
\ifx \oauthor \undefined \def \oauthor#1{#1}\fi
\ifx \citeauthoryear \undefined \def \citeauthoryear#1{#1}\fi
\ifx \endbibitem  \undefined \def \endbibitem {}\fi
\ifx \bconflocation  \undefined \def \bconflocation#1{#1}\fi
\ifx \arxivurl  \undefined \def \arxivurl#1{\textsf{#1}}\fi
\csname PreBibitemsHook\endcsname

%%% 1
\bibitem[\protect\citeauthoryear{United~Nations}{2019}]{united_nations_department_of_economic_and_social_affairs_population_division_world_2019}
\begin{bbook}
\bauthor{\bsnm{United~Nations}, \binits{P.D.} \bsuffix{Department of Economic {and} Social~Affairs}}:
\bbtitle{World Urbanization Prospects: the 2018 Revision}.
\bpublisher{United Nations},
\blocation{New York}
(\byear{2019}).
\bcomment{OCLC: 1120698127}
\end{bbook}
\endbibitem

%%% 2
\bibitem[\protect\citeauthoryear{Kowarik et~al.}{2025}]{kowarik_promoting_2025}
\begin{barticle}
\bauthor{\bsnm{Kowarik}, \binits{I.}},
\bauthor{\bsnm{Fischer}, \binits{L.K.}},
\bauthor{\bsnm{Haase}, \binits{D.}},
\bauthor{\bsnm{Kabisch}, \binits{N.}},
\bauthor{\bsnm{Kleinschroth}, \binits{F.}},
\bauthor{\bsnm{Konijnendijk}, \binits{C.}},
\bauthor{\bsnm{Straka}, \binits{T.M.}},
\bauthor{\bsnm{Haaren}, \binits{C.}}:
\batitle{Promoting urban biodiversity for the benefit of people and nature}.
\bjtitle{Nature Reviews Biodiversity}
\bvolume{1}(\bissue{4}),
\bfpage{214}--\blpage{232}
(\byear{2025})
\doiurl{10.1038/s44358-025-00035-y} .
\bcomment{Publisher: Nature Publishing Group}.
Accessed 2025-05-30
\end{barticle}
\endbibitem

%%% 3
\bibitem[\protect\citeauthoryear{Giannico et~al.}{2021}]{giannico_green_2021}
\begin{barticle}
\bauthor{\bsnm{Giannico}, \binits{V.}},
\bauthor{\bsnm{Spano}, \binits{G.}},
\bauthor{\bsnm{Elia}, \binits{M.}},
\bauthor{\bsnm{D’Este}, \binits{M.}},
\bauthor{\bsnm{Sanesi}, \binits{G.}},
\bauthor{\bsnm{Lafortezza}, \binits{R.}}:
\batitle{Green spaces, quality of life, and citizen perception in {European} cities}.
\bjtitle{Environmental Research}
\bvolume{196},
\bfpage{110922}
(\byear{2021})
\doiurl{10.1016/j.envres.2021.110922} .
Accessed 2024-07-31
\end{barticle}
\endbibitem

%%% 4
\bibitem[\protect\citeauthoryear{Browning et~al.}{2024}]{browning_measuring_2024}
\begin{barticle}
\bauthor{\bsnm{Browning}, \binits{M.H.E.M.}},
\bauthor{\bsnm{Locke}, \binits{D.H.}},
\bauthor{\bsnm{Konijnendijk}, \binits{C.}},
\bauthor{\bsnm{Labib}, \binits{S.M.}},
\bauthor{\bsnm{Rigolon}, \binits{A.}},
\bauthor{\bsnm{Yeager}, \binits{R.}},
\bauthor{\bsnm{Bardhan}, \binits{M.}},
\bauthor{\bsnm{Berland}, \binits{A.}},
\bauthor{\bsnm{Dadvand}, \binits{P.}},
\bauthor{\bsnm{Helbich}, \binits{M.}},
\bauthor{\bsnm{Li}, \binits{F.}},
\bauthor{\bsnm{Li}, \binits{H.}},
\bauthor{\bsnm{James}, \binits{P.}},
\bauthor{\bsnm{Klompmaker}, \binits{J.}},
\bauthor{\bsnm{Reuben}, \binits{A.}},
\bauthor{\bsnm{Roman}, \binits{L.A.}},
\bauthor{\bsnm{Tsai}, \binits{W.-L.}},
\bauthor{\bsnm{Patwary}, \binits{M.}},
\bauthor{\bsnm{O'Neil-Dunne}, \binits{J.}},
\bauthor{\bsnm{Ossola}, \binits{A.}},
\bauthor{\bsnm{Wang}, \binits{R.}},
\bauthor{\bsnm{Yang}, \binits{B.}},
\bauthor{\bsnm{Yi}, \binits{L.}},
\bauthor{\bsnm{Zhang}, \binits{J.}},
\bauthor{\bsnm{Nieuwenhuijsen}, \binits{M.}}:
\batitle{Measuring the 3-30-300 rule to help cities meet nature access thresholds}.
\bjtitle{Science of The Total Environment}
\bvolume{907},
\bfpage{167739}
(\byear{2024})
\doiurl{10.1016/j.scitotenv.2023.167739} .
Accessed 2024-12-09
\end{barticle}
\endbibitem

%%% 5
\bibitem[\protect\citeauthoryear{Zhang and Tan}{2019}]{zhang_associations_2019}
\begin{barticle}
\bauthor{\bsnm{Zhang}, \binits{L.}},
\bauthor{\bsnm{Tan}, \binits{P.Y.}}:
\batitle{Associations between {Urban} {Green} {Spaces} and {Health} are {Dependent} on the {Analytical} {Scale} and {How} {Urban} {Green} {Spaces} are {Measured}}.
\bjtitle{International Journal of Environmental Research and Public Health}
\bvolume{16}(\bissue{4}),
\bfpage{578}
(\byear{2019})
\doiurl{10.3390/ijerph16040578} .
Accessed 2025-07-02
\end{barticle}
\endbibitem

%%% 6
\bibitem[\protect\citeauthoryear{Battiston and Schifanella}{2024}]{battiston_need_2024}
\begin{barticle}
\bauthor{\bsnm{Battiston}, \binits{A.}},
\bauthor{\bsnm{Schifanella}, \binits{R.}}:
\batitle{On the need for a multi-dimensional framework to measure accessibility to urban green}.
\bjtitle{npj Urban Sustainability}
\bvolume{4}(\bissue{1}),
\bfpage{10}
(\byear{2024})
\doiurl{10.1038/s42949-024-00147-y} .
\bcomment{Publisher: Nature Publishing Group}.
Accessed 2025-07-02
\end{barticle}
\endbibitem

%%% 7
\bibitem[\protect\citeauthoryear{Juergens and Meyer-Heß}{2022}]{juergens_experimental_2022}
\begin{barticle}
\bauthor{\bsnm{Juergens}, \binits{C.}},
\bauthor{\bsnm{Meyer-Heß}, \binits{M.F.}}:
\batitle{Experimental {Analysis} of {Geo}-spatial {Data} to {Evaluate} {Urban} {Greenspace}: {A} {Case} {Study} in {Dortmund}, {Germany}}.
\bjtitle{KN - Journal of Cartography and Geographic Information}
\bvolume{72}(\bissue{2}),
\bfpage{153}--\blpage{171}
(\byear{2022})
\doiurl{10.1007/s42489-022-00107-5} .
Accessed 2025-01-30
\end{barticle}
\endbibitem

%%% 8
\bibitem[\protect\citeauthoryear{Boehnke et~al.}{2022}]{boehnke_mapping_2022}
\begin{barticle}
\bauthor{\bsnm{Boehnke}, \binits{D.}},
\bauthor{\bsnm{Krehl}, \binits{A.}},
\bauthor{\bsnm{Mörmann}, \binits{K.}},
\bauthor{\bsnm{Volk}, \binits{R.}},
\bauthor{\bsnm{Lützkendorf}, \binits{T.}},
\bauthor{\bsnm{Naber}, \binits{E.}},
\bauthor{\bsnm{Becker}, \binits{R.}},
\bauthor{\bsnm{Norra}, \binits{S.}}:
\batitle{Mapping {Urban} {Green} and {Its} {Ecosystem} {Services} at {Microscale}—{A} {Methodological} {Approach} for {Climate} {Adaptation} and {Biodiversity}}.
\bjtitle{Sustainability}
\bvolume{14}(\bissue{15}),
\bfpage{9029}
(\byear{2022})
\doiurl{10.3390/su14159029} .
\bcomment{Number: 15 Publisher: Multidisciplinary Digital Publishing Institute}.
Accessed 2025-01-30
\end{barticle}
\endbibitem

%%% 9
\bibitem[\protect\citeauthoryear{Derkzen et~al.}{2015}]{derkzen_review_2015}
\begin{barticle}
\bauthor{\bsnm{Derkzen}, \binits{M.L.}},
\bauthor{\bsnm{Teeffelen}, \binits{A.J.A.}},
\bauthor{\bsnm{Verburg}, \binits{P.H.}}:
\batitle{{REVIEW}: {Quantifying} urban ecosystem services based on high-resolution data of urban green space: an assessment for {Rotterdam}, the {Netherlands}}.
\bjtitle{Journal of Applied Ecology}
\bvolume{52}(\bissue{4}),
\bfpage{1020}--\blpage{1032}
(\byear{2015})
\doiurl{10.1111/1365-2664.12469} .
\bcomment{\_eprint: https://onlinelibrary.wiley.com/doi/pdf/10.1111/1365-2664.12469}.
Accessed 2025-01-30
\end{barticle}
\endbibitem

%%% 10
\bibitem[\protect\citeauthoryear{Seidl and Saifane}{2021}]{seidl_green_2021}
\begin{barticle}
\bauthor{\bsnm{Seidl}, \binits{M.}},
\bauthor{\bsnm{Saifane}, \binits{M.}}:
\batitle{A green intensity index to better assess the multiple functions of urban vegetation with an application to {Paris} metropolitan area}.
\bjtitle{Environment, Development and Sustainability}
\bvolume{23}(\bissue{10}),
\bfpage{15204}--\blpage{15224}
(\byear{2021})
\doiurl{10.1007/s10668-021-01293-4} .
Accessed 2025-01-30
\end{barticle}
\endbibitem

%%% 11
\bibitem[\protect\citeauthoryear{Kabisch et~al.}{2016}]{kabisch_urban_2016}
\begin{barticle}
\bauthor{\bsnm{Kabisch}, \binits{N.}},
\bauthor{\bsnm{Strohbach}, \binits{M.}},
\bauthor{\bsnm{Haase}, \binits{D.}},
\bauthor{\bsnm{Kronenberg}, \binits{J.}}:
\batitle{Urban green space availability in {European} cities}.
\bjtitle{Ecological Indicators}
\bvolume{70},
\bfpage{586}--\blpage{596}
(\byear{2016})
\doiurl{10.1016/j.ecolind.2016.02.029} .
Accessed 2024-12-12
\end{barticle}
\endbibitem

%%% 12
\bibitem[\protect\citeauthoryear{Robinson et~al.}{2022}]{robinson_urban_2022}
\begin{barticle}
\bauthor{\bsnm{Robinson}, \binits{J.M.}},
\bauthor{\bsnm{Mavoa}, \binits{S.}},
\bauthor{\bsnm{Robinson}, \binits{K.}},
\bauthor{\bsnm{Brindley}, \binits{P.}}:
\batitle{Urban centre green metrics in {Great} {Britain}: {A} geospatial and socioecological study}.
\bjtitle{PLOS ONE}
\bvolume{17}(\bissue{11}),
\bfpage{0276962}
(\byear{2022})
\doiurl{10.1371/journal.pone.0276962} .
\bcomment{Publisher: Public Library of Science}.
Accessed 2024-07-01
\end{barticle}
\endbibitem

%%% 13
\bibitem[\protect\citeauthoryear{Konijnendijk}{2023}]{konijnendijk_evidence-based_2023}
\begin{barticle}
\bauthor{\bsnm{Konijnendijk}, \binits{C.C.}}:
\batitle{Evidence-based guidelines for greener, healthier, more resilient neighbourhoods: {Introducing} the 3–30–300 rule}.
\bjtitle{Journal of Forestry Research}
\bvolume{34}(\bissue{3}),
\bfpage{821}--\blpage{830}
(\byear{2023})
\doiurl{10.1007/s11676-022-01523-z} .
Accessed 2024-07-31
\end{barticle}
\endbibitem

%%% 14
\bibitem[\protect\citeauthoryear{Zheng et~al.}{2024}]{zheng_quantitative_2024}
\begin{barticle}
\bauthor{\bsnm{Zheng}, \binits{Y.}},
\bauthor{\bsnm{Lin}, \binits{T.}},
\bauthor{\bsnm{Hamm}, \binits{N.A.S.}},
\bauthor{\bsnm{Liu}, \binits{J.}},
\bauthor{\bsnm{Zhou}, \binits{T.}},
\bauthor{\bsnm{Geng}, \binits{H.}},
\bauthor{\bsnm{Zhang}, \binits{J.}},
\bauthor{\bsnm{Ye}, \binits{H.}},
\bauthor{\bsnm{Zhang}, \binits{G.}},
\bauthor{\bsnm{Wang}, \binits{X.}},
\bauthor{\bsnm{Chen}, \binits{T.}}:
\batitle{Quantitative evaluation of urban green exposure and its impact on human health: {A} case study on the 3–30-300 green space rule}.
\bjtitle{Science of The Total Environment}
\bvolume{924},
\bfpage{171461}
(\byear{2024})
\doiurl{10.1016/j.scitotenv.2024.171461} .
Accessed 2024-06-20
\end{barticle}
\endbibitem

%%% 15
\bibitem[\protect\citeauthoryear{Nieuwenhuijsen et~al.}{2022}]{nieuwenhuijsen_evaluation_2022}
\begin{barticle}
\bauthor{\bsnm{Nieuwenhuijsen}, \binits{M.J.}},
\bauthor{\bsnm{Dadvand}, \binits{P.}},
\bauthor{\bsnm{Márquez}, \binits{S.}},
\bauthor{\bsnm{Bartoll}, \binits{X.}},
\bauthor{\bsnm{Barboza}, \binits{E.P.}},
\bauthor{\bsnm{Cirach}, \binits{M.}},
\bauthor{\bsnm{Borrell}, \binits{C.}},
\bauthor{\bsnm{Zijlema}, \binits{W.L.}}:
\batitle{The evaluation of the 3-30-300 green space rule and mental health}.
\bjtitle{Environmental Research}
\bvolume{215},
\bfpage{114387}
(\byear{2022})
\doiurl{10.1016/j.envres.2022.114387} .
Accessed 2024-06-20
\end{barticle}
\endbibitem

%%% 16
\bibitem[\protect\citeauthoryear{Croeser et~al.}{2024}]{croeser_acute_2024}
\begin{barticle}
\bauthor{\bsnm{Croeser}, \binits{T.}},
\bauthor{\bsnm{Sharma}, \binits{R.}},
\bauthor{\bsnm{Weisser}, \binits{W.W.}},
\bauthor{\bsnm{Bekessy}, \binits{S.A.}}:
\batitle{Acute canopy deficits in global cities exposed by the 3-30-300 benchmark for urban nature}.
\bjtitle{Nature Communications}
\bvolume{15}(\bissue{1}),
\bfpage{9333}
(\byear{2024})
\doiurl{10.1038/s41467-024-53402-2} .
\bcomment{Publisher: Nature Publishing Group}.
Accessed 2024-12-10
\end{barticle}
\endbibitem

%%% 17
\bibitem[\protect\citeauthoryear{Giannico et~al.}{2024}]{giannico_mortality_2024}
\begin{barticle}
\bauthor{\bsnm{Giannico}, \binits{O.V.}},
\bauthor{\bsnm{Sardone}, \binits{R.}},
\bauthor{\bsnm{Bisceglia}, \binits{L.}},
\bauthor{\bsnm{Addabbo}, \binits{F.}},
\bauthor{\bsnm{Pirotti}, \binits{F.}},
\bauthor{\bsnm{Minerba}, \binits{S.}},
\bauthor{\bsnm{Mincuzzi}, \binits{A.}}:
\batitle{The mortality impacts of greening {Italy}}.
\bjtitle{Nature Communications}
\bvolume{15}(\bissue{1}),
\bfpage{10452}
(\byear{2024})
\doiurl{10.1038/s41467-024-54388-7} .
\bcomment{Publisher: Nature Publishing Group}.
Accessed 2024-12-10
\end{barticle}
\endbibitem

%%% 18
\bibitem[\protect\citeauthoryear{Martin and Conway}{2025}]{martin_using_2025}
\begin{barticle}
\bauthor{\bsnm{Martin}, \binits{A.J.F.}},
\bauthor{\bsnm{Conway}, \binits{T.M.}}:
\batitle{Using the {Gini} {Index} to quantify urban green inequality: {A} systematic review and recommended reporting standards}.
\bjtitle{Landscape and Urban Planning}
\bvolume{254},
\bfpage{105231}
(\byear{2025})
\doiurl{10.1016/j.landurbplan.2024.105231} .
Accessed 2025-03-21
\end{barticle}
\endbibitem

%%% 19
\bibitem[\protect\citeauthoryear{Roussel et~al.}{2020}]{roussel_lidr_2020}
\begin{barticle}
\bauthor{\bsnm{Roussel}, \binits{J.-R.}},
\bauthor{\bsnm{Auty}, \binits{D.}},
\bauthor{\bsnm{Coops}, \binits{N.C.}},
\bauthor{\bsnm{Tompalski}, \binits{P.}},
\bauthor{\bsnm{Goodbody}, \binits{T.R.H.}},
\bauthor{\bsnm{Meador}, \binits{A.S.}},
\bauthor{\bsnm{Bourdon}, \binits{J.-F.}},
\bauthor{\bsnm{Boissieu}, \binits{F.}},
\bauthor{\bsnm{Achim}, \binits{A.}}:
\batitle{{lidR}: {An} {R} package for analysis of {Airborne} {Laser} {Scanning} ({ALS}) data}.
\bjtitle{Remote Sensing of Environment}
\bvolume{251},
\bfpage{112061}
(\byear{2020})
\doiurl{10.1016/j.rse.2020.112061} .
Accessed 2024-07-31
\end{barticle}
\endbibitem

%%% 20
\bibitem[\protect\citeauthoryear{Dalponte and Coomes}{2016}]{dalponte_tree-centric_2016}
\begin{barticle}
\bauthor{\bsnm{Dalponte}, \binits{M.}},
\bauthor{\bsnm{Coomes}, \binits{D.A.}}:
\batitle{Tree-centric mapping of forest carbon density from airborne laser scanning and hyperspectral data}.
\bjtitle{Methods in Ecology and Evolution}
\bvolume{7}(\bissue{10}),
\bfpage{1236}--\blpage{1245}
(\byear{2016})
\doiurl{10.1111/2041-210X.12575} .
\bcomment{\_eprint: https://onlinelibrary.wiley.com/doi/pdf/10.1111/2041-210X.12575}.
Accessed 2024-07-31
\end{barticle}
\endbibitem

%%% 21
\bibitem[\protect\citeauthoryear{Montero et~al.}{2023}]{montero_standardized_2023}
\begin{barticle}
\bauthor{\bsnm{Montero}, \binits{D.}},
\bauthor{\bsnm{Aybar}, \binits{C.}},
\bauthor{\bsnm{Mahecha}, \binits{M.D.}},
\bauthor{\bsnm{Martinuzzi}, \binits{F.}},
\bauthor{\bsnm{Söchting}, \binits{M.}},
\bauthor{\bsnm{Wieneke}, \binits{S.}}:
\batitle{A standardized catalogue of spectral indices to advance the use of remote sensing in {Earth} system research}.
\bjtitle{Scientific Data}
\bvolume{10}(\bissue{1}),
\bfpage{197}
(\byear{2023})
\doiurl{10.1038/s41597-023-02096-0} .
\bcomment{Publisher: Nature Publishing Group}.
Accessed 2025-07-31
\end{barticle}
\endbibitem

%%% 22
\bibitem[\protect\citeauthoryear{Zha et~al.}{2003}]{zha_use_2003}
\begin{barticle}
\bauthor{\bsnm{Zha}, \binits{Y.}},
\bauthor{\bsnm{Gao}, \binits{J.}},
\bauthor{\bsnm{Ni}, \binits{S.}}:
\batitle{Use of normalized difference built-up index in automatically mapping urban areas from {TM} imagery}.
\bjtitle{International Journal of Remote Sensing}
\bvolume{24}(\bissue{3}),
\bfpage{583}--\blpage{594}
(\byear{2003})
\doiurl{10.1080/01431160304987} .
\bcomment{Publisher: Taylor \& Francis \_eprint: https://doi.org/10.1080/01431160304987}.
Accessed 2025-07-04
\end{barticle}
\endbibitem

%%% 23
\bibitem[\protect\citeauthoryear{Zheng et~al.}{2021}]{zheng_improved_2021}
\begin{barticle}
\bauthor{\bsnm{Zheng}, \binits{Y.}},
\bauthor{\bsnm{Tang}, \binits{L.}},
\bauthor{\bsnm{Wang}, \binits{H.}}:
\batitle{An improved approach for monitoring urban built-up areas by combining {NPP}-{VIIRS} nighttime light, {NDVI}, {NDWI}, and {NDBI}}.
\bjtitle{Journal of Cleaner Production}
\bvolume{328},
\bfpage{129488}
(\byear{2021})
\doiurl{10.1016/j.jclepro.2021.129488} .
Accessed 2025-07-04
\end{barticle}
\endbibitem

%%% 24
\bibitem[\protect\citeauthoryear{Martinez and Labib}{2023}]{martinez_demystifying_2023}
\begin{barticle}
\bauthor{\bsnm{Martinez}, \binits{A.d.l.I.}},
\bauthor{\bsnm{Labib}, \binits{S.M.}}:
\batitle{Demystifying normalized difference vegetation index ({NDVI}) for greenness exposure assessments and policy interventions in urban greening}.
\bjtitle{Environmental Research}
\bvolume{220},
\bfpage{115155}
(\byear{2023})
\doiurl{10.1016/j.envres.2022.115155} .
Accessed 2025-07-04
\end{barticle}
\endbibitem

%%% 25
\bibitem[\protect\citeauthoryear{Li et~al.}{2015}]{li_comparison_2015}
\begin{barticle}
\bauthor{\bsnm{Li}, \binits{W.}},
\bauthor{\bsnm{Saphores}, \binits{J.-D.M.}},
\bauthor{\bsnm{Gillespie}, \binits{T.W.}}:
\batitle{A comparison of the economic benefits of urban green spaces estimated with {NDVI} and with high-resolution land cover data}.
\bjtitle{Landscape and Urban Planning}
\bvolume{133},
\bfpage{105}--\blpage{117}
(\byear{2015})
\doiurl{10.1016/j.landurbplan.2014.09.013} .
Accessed 2025-07-04
\end{barticle}
\endbibitem

%%% 26
\bibitem[\protect\citeauthoryear{Yang et~al.}{2017}]{yang_mapping_2017}
\begin{barticle}
\bauthor{\bsnm{Yang}, \binits{X.}},
\bauthor{\bsnm{Zhao}, \binits{S.}},
\bauthor{\bsnm{Qin}, \binits{X.}},
\bauthor{\bsnm{Zhao}, \binits{N.}},
\bauthor{\bsnm{Liang}, \binits{L.}}:
\batitle{Mapping of {Urban} {Surface} {Water} {Bodies} from {Sentinel}-2 {MSI} {Imagery} at 10 m {Resolution} via {NDWI}-{Based} {Image} {Sharpening}}.
\bjtitle{Remote Sensing}
\bvolume{9}(\bissue{6}),
\bfpage{596}
(\byear{2017})
\doiurl{10.3390/rs9060596} .
\bcomment{Number: 6 Publisher: Multidisciplinary Digital Publishing Institute}.
Accessed 2025-07-04
\end{barticle}
\endbibitem

\end{thebibliography}


\end{document}